\documentclass[%
 pre,
 amsmath,amssymb,
 reprint,%
]{revtex4-1}

\usepackage{graphicx}
\usepackage{dcolumn}
\usepackage{bm}

\usepackage[utf8]{inputenc}
\usepackage[T1]{fontenc}
\usepackage{mathptmx}
\usepackage{hyperref}
\usepackage{etoolbox}
\makeatletter
\def\@email#1#2{%
 \endgroup
 \patchcmd{\titleblock@produce}
  {\frontmatter@RRAPformat}
  {\frontmatter@RRAPformat{\produce@RRAP{*#1\href{mailto:#2}{#2}}}\frontmatter@RRAPformat}
  {}{}
}%
\DeclareUnicodeCharacter{00A0}{ }
\makeatother
\begin{document}

\preprint{AIP/123-QED}

\title[]{Microscopic Activated Dynamics Theory of Shear Rheology and the Stress Overshoot in Ultra-Dense Glass-Forming Fluids and Colloidal Suspensions}

\author{Ashesh Ghosh}
\altaffiliation{Present address: Department of Chemical Engineering, Stanford University, CA 64305}
\affiliation{
 Department of Chemistry, University of Illinois at Urbana-Champaign, IL 61801 USA
 }
\affiliation{
 Materials Research Laboratory, University of Illinois at Urbana-Champaign, IL 61801 USA
 }
\author{Kenneth S. Schweizer}%
 \email{kschweiz@illinois.edu.}
 \affiliation{
 Department of Chemistry, University of Illinois at Urbana-Champaign, IL 61801 USA
 }
 \affiliation{
 Materials Research Laboratory, University of Illinois at Urbana-Champaign, IL 61801 USA
 }
 \affiliation{
 Department of Materials Science, University of Illinois at Urbana-Champaign, IL 61801 USA
 }
 \affiliation{
 Department of Chemical \& Biomolecular Engineering, University of Illinois at Urbana-Champaign, IL 61801 USA
 }
 \affiliation{
 Beckman Institute, University of Illinois at Urbana-Champaign, IL 61801 USA
 }

\date{\today}

\begin{abstract}
We formulate a microscopic, force-level, activated dynamics-based theory for the continuous startup shear rheology of ultra-dense glass-forming hard sphere fluids and colloidal suspensions in the context of elastically collective nonlinear Langevin equation (ECNLE) theory. 
This microscopic force-based approach describes activated relaxation as a coupled local–nonlocal event involving caging and longer-range collective elasticity. 
ECNLE based theoretical predictions for the deformation-induced enhancement of mobility, the onset of relaxation speed up at remarkably low values of stress, strain, or shear rate, apparent power law thinning of the steady state structural relaxation time and viscosity, a non-vanishing activation barrier in the shear thinning regime, an apparent Herschel–Bulkley form of the rate dependence of the steady state shear stress, exponential growth of different measures of a dynamic yield or flow stress with the packing fraction, and reduced fragility and dynamic heterogeneity under deformation were shown to be in excellent agreement with experimental results. 
The central new question we address is that of stress overshoot in transient shear rheology. 
In contrast to the steady state flow regime, understanding the complex transient response requires an explicit treatment of the coupled nonequilibrium evolution of structure, elastic modulus and stress relaxation time. 
Theoretical predictions for the stress overshoot are shown to be in good agreement with experimental observations as a function of shear rate and packing fraction in all rheological regimes, and deformation-assisted activated motion is shown to be crucial for both the transient and steady state responses. 
\end{abstract}

\maketitle

\section{Introduction}
Microscopic understanding of the nonlinear mechanical response of deeply supercooled or over-compressed liquids and glasses under strong deformation is a fundamental problem (with high technological relevance) in nonequilibrium statistical mechanics for diverse systems including colloids, metals and polymers ~\cite{mewis2012colloidal,hebert2015effect,berthier2011theoretical,gratson2004direct,lee2009direct}. 
Quiescent structural relaxation involves thermally activated hopping processes which can be massively sped under deformation, ultimately resulting in (plastic) flow, in a manner that depends on thermodynamic state, interparticle interactions, and deformation rate~\cite{lee2009direct,eyring1936viscosity,jadhao2017probing,pham2006yielding,besseling2007three,lee2009deformation,koumakis2012direct}. 
For the classic rheological protocol of homogeneous continuous startup shear, the stress evolution exhibits multiple regimes with increasing time or accumulated strain (see, e.g., Fig.1a): (i) an initial linear response characterized by an elastic shear modulus (blue shaded region in Fig. 1(a)), (ii) an ‘anelastic’ (or, nonlinear elastic) regime where dissipative processes begin to emerge (yellow shaded region in Fig. 1(a)), (iii) a stress maximum or “overshoot” (point iii in Fig. 1(a)), (iv) stress softening (gray shaded region in Fig. 1(a)), and (v) steady state flow characterized by a plateau stress which determines the nonequilibrium viscosity (red shaded region in Fig. 1(a))~\cite{larson1999structure,haward1997physics,roth2016polymer,chen2011theory,koumakis2016a,koumakis2012a,laurati2012a}.

Of special interest is the physics of the defining feature of a non-monotonic stress-strain curve, the overshoot. 
This stress maximum is often intuitively viewed as a “dynamic yield point” that signals a mechanically driven crossover from predominantly elastic to predominantly viscous response. 
For the ultra-dense monodisperse hard sphere fluid and Brownian metastable colloidal suspensions of present interest which have a well-defined maximum or random close packing (RCP) fraction, the overshoot strain and stress are rich functions of shear rate and concentration. 
This includes a striking non-monotonic evolution of the overshoot amplitude with shear rate, and its tendency to vanish at ultra-high packing fractions~\cite{koumakis2016a,koumakis2012a}. 
The microscopic physics of these observations are not understood, and seem especially subtle since the overshoot amplitude reflects the relative magnitudes of the transient peak versus steady state stresses, and hence can potentially involve different nonequilibrium processes. We believe that at least three physical effects are crucial: (i) activated structural relaxation under quiescent and strongly deformed conditions as objectively indicated, e.g., by the observation in experiments and simulations of intermittent particle trajectories even far from equilibrium at high stresses~\cite{besseling2007three}, (ii) accelerated relaxation via shear advection and/or stress reduction of activation barriers [6,19-21], and (iii) deformation-induced change of structure which typically weakens (in an average sense) cage scale spatial correlations and kinetic constraints [12,16-18].  

\begin{figure*}[ht]
    \centering
    \includegraphics[width=0.95\textwidth]{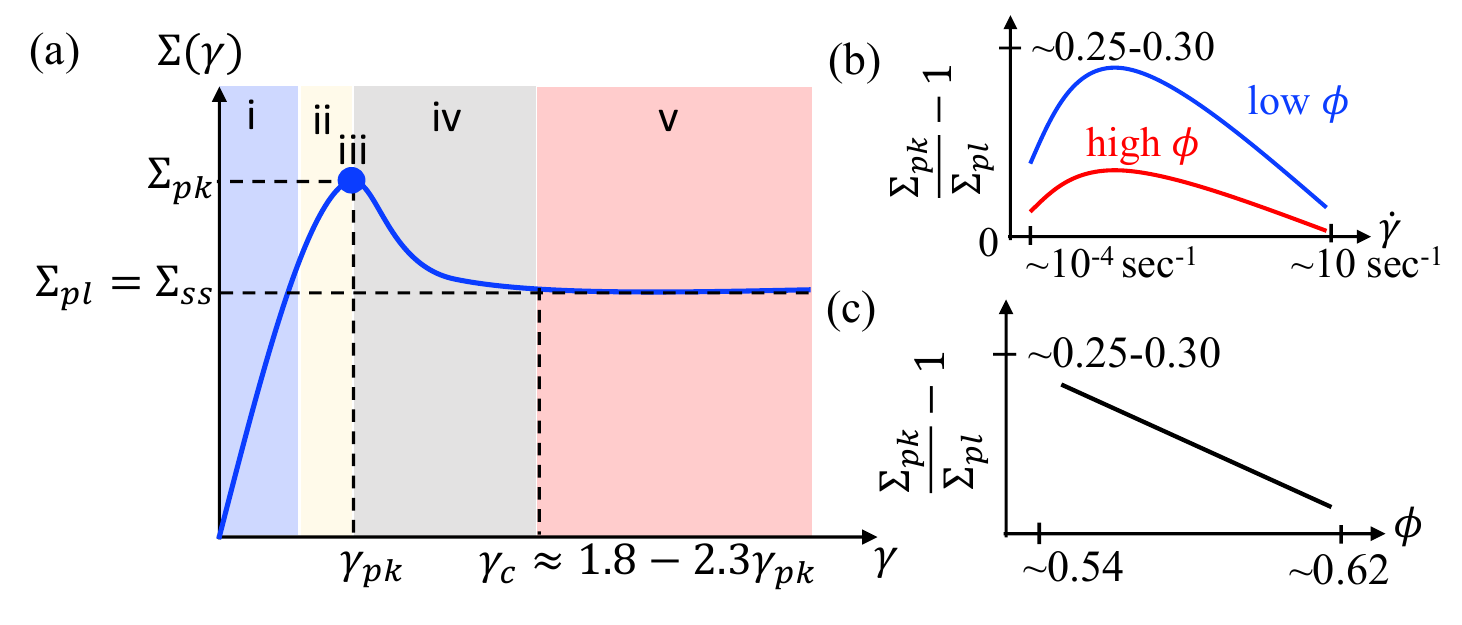}
    \caption{Schematic of (a) the stress ($\Sigma$) vs strain ($\gamma$) curve for a fixed strain rate that shows the various regions of behavior. 
    A stress overshoot is observed at a strain of $\gamma=\gamma_{\text{pk}}$, which increases weakly with shear rate. The overshoot stress and plateau (or steady state) stress are indicated as $\Sigma_{\text{pk}}$ and $\Sigma_{\text{ss}}$ (or $\Sigma_{\text{pl}}$), respectively.
    For dense hard sphere colloidal suspensions, experiments and simulations find the emergence of a non-equilibrium steady state typically occurs at a strain 1.8-2.3 times larger than the overshoot peak strain. 
    Schematic of the amplitude of the stress overshoot, defined as, $r_O=\Sigma_{\text{pk}}/\Sigma_{\text{pl}}-1$, as a function of strain rate at fixed low and high packing fractions in the dense metastable (``glassy'') regime are shown in (b) which displays a characteristic non-monotonic behavior. 
    (c) Monotonically decreasing overshoot amplitude as a function of packing fraction at a fixed strain rate.}
    \label{fig:sch_overshoot}
\end{figure*}

Conceptually diverse phenomenological plasticity models (e.g., soft glass rheology (SGR), shear-transformation zone (STZ), mean-field nonlinear deformation) have been developed for the rheology of non-polymeric systems~\cite{falk1998a,demetriou2006a,fuchs2002a,zaccone2020a,zaccone2014a}. 
STZ models invoke specific average features (such as the population density STZs) of a relatively compact two-state shear-transformation zone with internal dynamics to understand the large-scale deformation behavior of amorphous solids~\cite{langer2006a}. 
SGR models often ignore tensorial aspects of deformation for simplicity and construct a probabilistic evolution equation of a specified region within a material as a function of strain, strain rate, a yield stress, and time. 
More recently, stress overshoots in simple yield stress fluids have been analyzed by a “continuum fluidity” model~\cite{benzi2021a,benzi2021b} that combines a fluidity equation with stress evolution per a Maxwell model. 
However, all these models are phenomenological in the sense that the 3 points enumerated above are not microscopically addressed and/or empirical fit parameter(s) are invoked. 
Moreover, they generally are applied to soft matter suspensions (e.g., microgels, pastes,) that do not exhibit a literal RCP state, but rather a soft jamming crossover.  
More crucially for our interests, the 5 zones of the stress-strain curve in Fig. 1a as a function of strain rate and packing fraction are not predicted based on a microscopic description.

The most ambitious microscopic “first principles” approach formulated at the level of forces and static and dynamic structural correlations is the mode coupling theory (MCT)~\cite{fuchs2002a,amann2013a,fuchs2002b,fuchs2005a,amann2015a,goetze2008a}.
The key physics for nonlinear rheology is argued to emerge from local structural changes driven by shear advection that weaken and ultimately destroy local caging constraints~\cite{fuchs2002a,amann2013a,fuchs2002b,fuchs2005a,amann2015a}. 
Though valuable progress has been made with this approach, qualitative problems remain. 
These include the prediction of a strong increase of the overshoot amplitude with packing fraction over a tiny change of colloid concentration and a monotonic evolution of the overshoot amplitude with shear rate~\cite{amann2015a}, both in disagreement with experiment and simulation~\cite{koumakis2016a,koumakis2012a,laurati2012a} (see Fig. 13 in Appendix). 
A major conceptual issue is that the rheological MCT is built on a hypothetical quiescent ``ideal glass transition''~\cite{goetze2008a}, the existence of which reflects the neglect of ergodicity restoring thermally activated processes. 
The numerical complexity of the fully microscopic rheological MCT has also largely restricted its applications~\cite{amann2015a} to further simplified semi-phenomenological “schematic” versions which contain multiple fit parameters, and in some cases a fit function. 
The version of ideal MCT for sheared colloidal suspensions in ref.~\cite{miyazaki2002a} utilizes a shear distorted structure factor and a generalized hydrodynamics approach where shearing retains the form of the force-force memory kernel in MCT with modified time-dependent wave vector associated with advection. 
However, this approach is restricted to 2D fluids due to the large numerical complexity associated with the tensorial nature of the structure factor.

Here we take a different microscopic theoretical approach to address the nonlinear rheology of ultra-dense metastable hard sphere systems by building on the Elastically Collective Nonlinear Langevin Equation (ECNLE) theory~\cite{mirigian2013a,mirigian2014a} of coupled local cage-nonlocal collective elastic activated structural relaxation in equilibrium glass-forming fluids (conceptual schematic in Fig. 2a), and its microrheology-inspired generalization to treat deformation~\cite{ghosh2020a}. 
The central quantity is a dynamic free energy, $F_{\text{dyn}} (r)$, that is a function of instantaneous scalar particle displacement (r) and which controls single particle stochastic trajectories including thermal activation effects based on a nonlinear Langevin equation (NLE) of motion. 
The coupling of cage scale barrier hopping and longer range collective elasticity of the fluid is included, with the latter becoming increasingly important in equilibrium as the fluid packing fraction (temperature in a viscous liquid) becomes very high (low). 
In the spirit of MCT, the dynamic free energy quantifies kinetic constraints via the pair correlation function ($g(r)$ or Fourier space structure factor $S(k)$). 
In prior work to date, stress enters the NLE evolution equation solely as an external force which directly accelerates relaxation via softening of the constraints embedded in the nonequilibrium free energy, with stress-induced activation barrier reduction being the most important aspect.

The formulation of ECNLE theory sketched above has previously addressed points (i) and (ii) discussed at the start of this Introduction, but not point (iii) since S(k) was assumed to be invariant to deformation. 
This zeroth order assumption of an invariant local structure model is of course not literally true, but has some practical support from measurements on of model silica hard sphere suspensions~\cite{watanabe1997a,watanabe1998a,maranzano2002a} where, (a) no significant changes of structure are observed up to a packing fraction of 0.50, and (b) surprisingly, strong nonlinear dynamical effects are not accompanied by significant anisotropy of structure in the shear thinning regime~\cite{maranzano2002a}. 
The invariant-structure theory~\cite{ghosh2020a}  predicts massive deformation-induced mobility enhancement leading to the following predictions that have been shown to agree well with experiments and simulations: (a) nonlinearity emerges at remarkably low values of “renormalized” Peclet number (product of shear rate and quiescent structural or alpha relaxation time, $\text{Pe}=\dot{\gamma}\tau_\alpha\ll 1$), (b) power law alpha time and viscosity thinning in the steady state, (c) a flow curve of the Herschel-Buckley (HB) empirical form~\cite{larson1999structure}, and (d) an apparent static yield stress that grows nearly exponentially with packing fraction~\cite{ghosh2020a}. 
More broadly, we note that recent combined confocal imaging and rheology experiments~\cite{sentjabrskaja2015a} provide significant justification for our specific theoretical formulation of rheological response built on single dynamics since a very strong link between macroscopic nonlinear creep with microscopic single particle motion as encoded at the elementary mean square displacement level has been observed.
However, the full transient stress-strain response, the stress overshoot, and the fundamental problem of dynamic yielding have not been addressed.
Doing so is the goal of the present article. 
We formulate a minimalist microscopic model for capturing the coupled nonequilibrium evolution of structure, relaxation time, and rheology which is demonstrated to provide a unified understanding of the stress overshoot, the steady state, and the role of activated dynamics.

\section{ECNLE Theory}

ECNLE theory under both quiescent and driven conditions with a deformation-invariant $S(k)$ has been thoroughly described in the literature~\cite{mirigian2013a,mirigian2014a,ghosh2020a,kobelev2005a}. 
For the convenience of the reader, and as relevant background, we briefly review without derivation the key elements germane to our new generalization (entering solely via the deformation-dependent isotropic structure factor, $S(k)$) in the context of the system of present interest, a fluid or suspension of monodisperse hard spheres.
The key input to the quiescent dynamical theory is $S(k)=1+\rho h(k) = (1-\rho C(k))^{-1}$, where $h(k)$ is the Fourier transform of the non-random part of the pair correlation function, $h(r)=g(r)-1$, and $C(k)$ is the direct correlation function~\cite{hansen2006a}, computed here using standard PY integral equation theory~\cite{hansen2006a}.

\begin{figure*}[ht]
    \centering
    \includegraphics[width=0.95\textwidth]{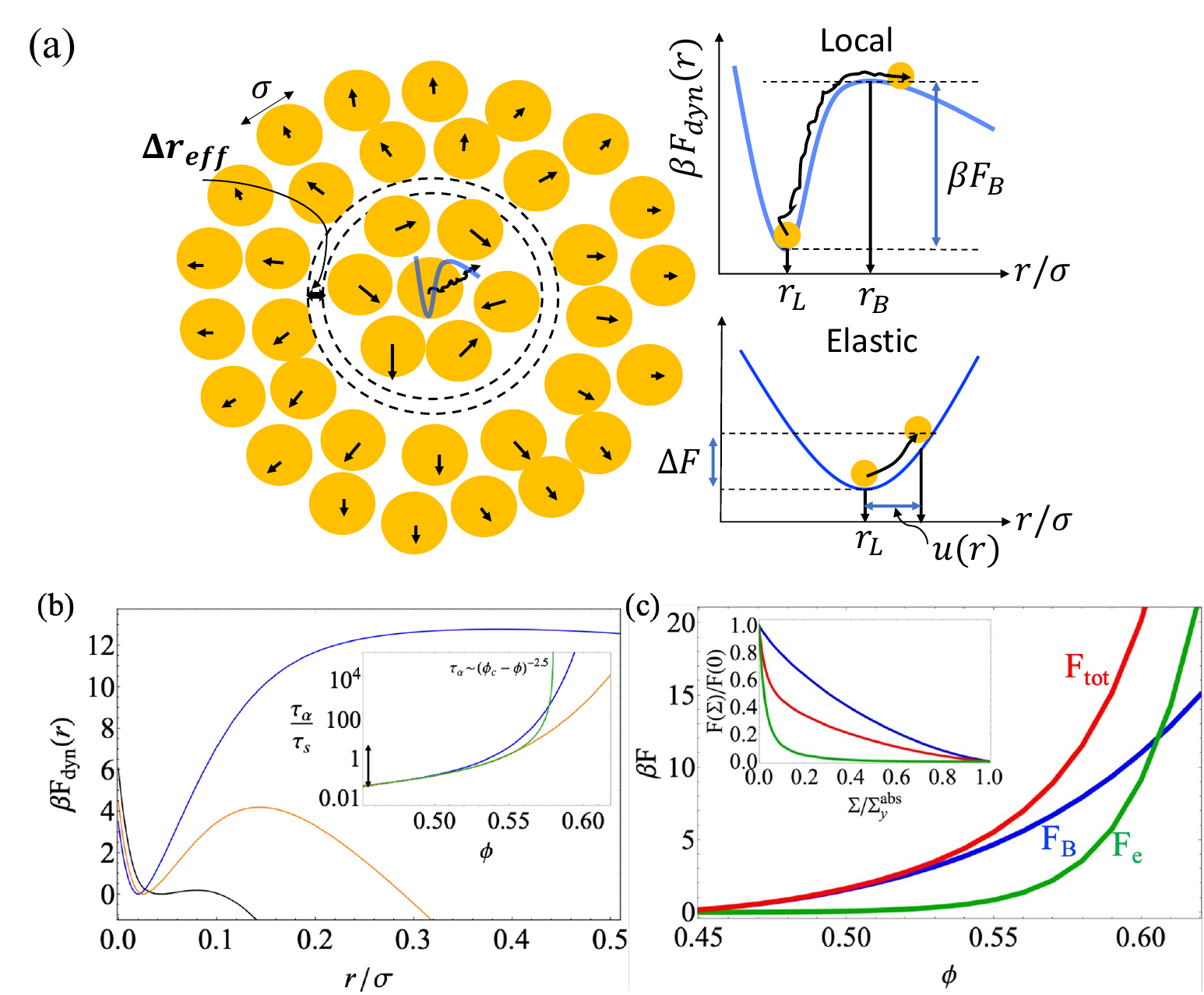}
    \caption{(a) Schematic of the activated hopping of particles on the cage scale described by NLE theory which is coupled to a longer ranged collective elastic barrier arising from the effective cage expansion and displacement of all particles outside the cage. Schematic of the NLE dynamic free energy is shown with all important length scales indicated (top right), while the collective elastic barrier arises from the small harmonic displacements of all particles outside the cage (bottom right). (b) Evolution of the dynamic free energy as a function of particle displacement under a constant applied stress for a high packing fraction of 0.61. Blue, orange and black curves are for, respectively, zero stress, an intermediate stress, and the ‘absolute yield stress’ ($\Sigma_y^{\text{abs}}$) defined as the value of stress required to first destroy the cage barrier ($\beta F_B=0$)~\cite{koumakis2016a,koumakis2012a} and the localized form of the dynamic free energy. The inset shows the equilibrium alpha time normalized by the short process time scale as a function of packing fraction predicted by NLE theory (orange; no collective elasticity) and ECNLE theory (blue). The green curve is the horizontally and vertically shifted ideal MCT prediction based on an empirically shifted packing fraction $\phi_c\sim 0.581$, which agrees with ECNLE theory over the first few decades of relaxation indicated by a vertical black arrow. (c) Plot of the local (blue), elastic (green) and total (red) barriers as a function of packing fraction. Inset: the normalized changes of the 3 barriers with stress nondimensionalized by the absolute yield stress for a packing fraction of 0.61.}
    \label{fig:sch_ecnle}
\end{figure*}
\subsection{Quiescent fluids}
The dynamic starting point is the nonlinear Langevin equation (NLE) that describes the stochastic evolution of a tagged particle trajectory controlled by a dynamic free energy. 
The dynamical variable is the angularly averaged scalar displacement of a particle, $r(t)$, which under quiescent conditions obeys the stochastic NLE in the overdamped limit~\cite{schweizer2003a,schweizer2005a}:
\begin{equation}
    -\zeta_s\frac{dr(t)}{dt} -\frac{\partial F_{\text{dyn}(r(t))}}{\partial t} +\xi(t) = 0
\end{equation}
where $\xi(t)$  is a white noise random force associated with a short time Fickian diffusion process quantified by the friction constant $\zeta_s$ determined based on a binary collision plus weak caging model known to be accurate for hard sphere colloidal suspensions when trajectories are not strongly activated~\cite{verberg1997a}; hydrodynamics enters only via the single particle Stokes-Einstein diffusivity. 
An explicit expression for $\zeta_s$ is given below.  
The dynamic free energy is~\cite{schweizer2003a,schweizer2005a},
\begin{equation}
\begin{split}
     \beta F_{\text{dyn}(r)} = -3\ln{\left(\frac{r}{\sigma}\right)} - \rho\int\frac{d\vec{k}}{(2\pi)^3} \frac{C(k)^2S(k)}{\left(1+S^{-1}(k)\right)}\times \\
    \exp{\left(-\frac{k^2r^2}{6}(1+S^{-1}(k))\right)}
\end{split}
\end{equation}
where the integrand contains the spatially resolved (in Fourier space) mean square force vertex defined as $V(k)=\rho k^2 C(k)^2 S(k)$ that quantifies caging constraints on a length scale $2\pi/k$ per the MCT idea of projecting real forces onto bilinear density variables followed by a Gaussian factorization approximation of 4-point correlation to products of pair correlation functions~\cite{wolynes2012a,kirkpatrick1985a}. 
The so-called ideal MCT “transition” at the single particle level (referred to as naïve MCT or, NMCT) is recovered from the mathematical condition of when a local minimum of  the dynamic free energy first emerges with increasing packing fraction~\cite{schweizer2003a,schweizer2005a}. 
This connection between NMCT and NLE theory is equivalent to stating NMCT effectively ignores the thermal noise in the NLE evolution equation, i.e. thermal fluctuation driven “uphill” activated motions are not allowed. 
The emergence of transient localization length ($r_L$), defined as the local minimum of the dynamic free energy, at $\phi=\phi_c=0.432$ (using PY closure~\cite{hansen2006a}) defines a smooth dynamic crossover to the activated regime in NLE theory.
For $\phi>\phi_c$ the dynamic free energy has a local cage barrier of height $F_B$ at $r=r_B$, with a jump distance defined naturally as $\Delta r=r_B-r_L$. 
Fig. 1a shows an example of the dynamic free energy in a highly activated regime where $\phi=0.61$. 
The local cage barrier is shown in Fig. 2c as a function of packing fraction.

NLE theory has been shown to capture only noncooperative hopping events and hence is valid only at sufficiently low packing fractions (but still in the overcompressed or metastable $\phi >0.495$ regime for monodisperse hard spheres)~\cite{mirigian2013a,mirigian2014a}. 
This motivated the extension of NLE theory (thereby defining ECNLE theory) to include collective physics associated with the coupling of local hopping on the cage scale and collective elasticity associated with the facilitating displacement of all particles beyond the cage size. 
The basic physical idea was inspired by the phenomenological shoving model of Dyre~\cite{dyre1998a,dyre2006a} (albeit with multiple physical differences as discussed elsewhere~\cite{mirigian2014a}) that all particles outside the cage must elastically displace via a spontaneous fluctuation to create the small amount of extra space required to allow large amplitude cage scale particle hopping~\cite{mirigian2013a,mirigian2014a}. 
The corresponding elastic energy cost enters as an additional barrier for relaxation which has been derived to be~\cite{mirigian2014a},

\begin{equation}
    \begin{split}
        \beta F_e = 4\pi\int_{r_{\text{cage}}}^\infty dr r^2 \rho g(r) \left(\frac{1}{2}K_0 u(r)^2\right) \\
        \approx 12\phi K_0 \left(\frac{r_{\text{cage}}}{\sigma}\right)^3 \Delta r_{\text{eff}}^2
    \end{split}
\end{equation}

where $K_0=3k_BT/r_L^2$ is the harmonic spring constant or curvature of the dynamic free energy at its local minimum;$\Delta r_{\text{eff}}=\frac{3}{32}\frac{\Delta r^2}{r_{\text{cage}}}$ is the effective amplitude (typically small, of order $r_L$ or smaller) of the elastic displacement field at the cage surface. 
The elastic displacement field at larger scalar distances $r$ from the cage center is $u(r)=\Delta r_{\text{eff}} (r_{\text{cage}}/r)^2$,  $r_{\text{cage}}$ is the mean cage radius defined by the first minimum of $g(r)$ ($\sim 1.5\sigma$ for hard spheres) and the final relation in Eq(3) uses $g(r)=1$ for $r\ge r_{\text{cage}}$ which is a technical simplification that has been shown to be extremely accurate for hard spheres~\cite{mirigian2014a}. 
In ECNLE theory, the alpha relaxation event is a coupled local-nonlocal process corresponding to a total barrier equal to the sum of the local cage and collective elastic contributions, $\beta F_\text{{total}}=\beta (F_B+F_e)$. 
Importantly, all quantities needed to compute the total barrier follow from the dynamic free energy, and hence the structural correlations. The average hopping or alpha relaxation time is computed using a Kramers mean first passage time analysis for barrier crossing as~\cite{mirigian2014a},

\begin{equation*}
    \begin{split}
        \frac{\tau_\alpha}{\tau_s} = \frac{2}{\sigma^2}\exp{(\beta F_e)}\int_{r_L}^{r_B} dx \exp{(\beta F_{\text{dyn}}(x))}\times \\ \int_{r_L}^{x} dy \exp{(-\beta F_{\text{dyn}}(y))} \\
        \approx \frac{2\pi}{\sqrt{K_0K_B}}\exp{(\beta F_{\text{total}})}
    \end{split}
    \tag{4a}
\end{equation*}
where $\tau_s=\beta \zeta_s \sigma^2$ is a short relaxation timescale, here taken as the elementary time unit, and for hard spheres is given by~\cite{cohen1998a}
\begin{equation*}
    \frac{\tau_s}{\tau_0} = 1+\frac{\sigma^3}{36\pi\phi}\int_0^{\infty} dk \left( \frac{k^2(S(k)-1)^2}{S(k)+(1-j_0(k\sigma)+2j_2(k\sigma))^{-1}}\right)
    \tag{4b}
\end{equation*}

where $j_n (x)$ is spherical Bessel function of order n and the ‘bare’ relaxation time $\tau_0$ can be written in terms of Stokes-Einstein friction constant ($\zeta_{\text{SE}}$) for suspensions with a binary collision contribution as, $\tau_0=\beta g(\sigma)\sigma^2 \zeta_{\text{SE}}$. 
The simpler approximate Kramers result in Eq. (4a) is valid when the local barrier is $\sim 1-2 k_B T$ or higher.

A sample calculation of the equilibrium alpha time with and without the elastic barrier is shown in the inset of Fig. 2b. 
The elastic barrier is not important for lower packing fractions in the metastable regime where the barrier is determined by local caging, but it becomes increasingly dominant for $\phi\ge 0.56$. 
The critical power law form of ideal MCT (appropriately shifted) is shown as the green curve in the inset of Fig. 2b. 
As discussed previously~\cite{schweizer2003a}, this critical power law form empirically captures the initial $\sim 2-2.5$ decades of slowdown (low barrier regime) predicted by the activated dynamics theory before it unphysically diverges as $\tau_\alpha\sim (\phi_{\text{MCT}}^{\text{shift}}-\phi)^{-2.5}$.

\subsection{Effect of external forces}

Motivated by a microrheology perspective in a stress ensemble, the direct effect of external deformation in the NLE approach enters as an additional external force, $f$, on a tagged particle corresponding to a nonequilibrium dynamic free energy with an extra contribution of a mechanical work form~\cite{ghosh2020a,kobelev2005a}:
\begin{equation}
    \beta F_{\text{dyn}}(r,\Sigma) = \beta F_{\text{dyn}}(r,\Sigma=0) - f\cdot r
    \tag{5}
\end{equation}
where $\Sigma$ is stress. A linear connection between microscopic force and macroscopic stress is employed: $f=A\Sigma$, where $A$ is a microscopic cross-sectional area related to particle size. 
A detailed discussion of the motivation for Eq(5) and the specific choice of $A$ is provided in refs. ~\cite{ghosh2020a,kobelev2005a}; here we briefly explain the main ideas. 

For computational tractability and conceptual simplicity reasons, the NLE theory is formulated based on scalar displacement of particles. 
Hence, prior rheological work with this approach has adopted an isotropized description of the dynamic trajectories under deformation for which treating $S(k)$ as isotropic is consistent.
This is why only a scalar force, $f$, enters Eq(5), which can be viewed as a pre-averaging of the anisotropy (different directions of external force and particle trajectory) for which one obtains $A=\pi\sigma^2/24$~\cite{ghosh2020a}; another choice is the particle cross-sectional area of $A=\pi\sigma^2/4$. 
These 2 choices differ only by a numerical prefactor and we have verified here and elsewhere~\cite{ghosh2020a} that our results are insensitive to the precise choice which only affects the absolute values of characteristic rheological quantities but not the basic qualitative behaviors predicted by the theory~\cite{ghosh2020a}. 
For the rest of the article, we adopt $A=\pi\sigma^2/24$. 

The mean alpha time is again computed using Kramers theory but with all required properties now stress-dependent, except for the very short time and length scale process prefactor $\tau_s=\beta\zeta_s\sigma^2$ in Eq. 4a which is fixed at its equilibrium value. 
With increasing applied force at a fixed packing fraction, one physically expects (as Eq (5) predicts) that  $r_L$ increases indicating weaker localization. 
The ‘critical’ force or stress at which the localized form of the dynamic free energy first vanishes, the local barrier goes to zero continuously and the localization length disappears, is termed the ‘absolute yield stress’, $\Sigma_y^{\text{abs}}$. 
Both the local and elastic barriers decrease with applied deformation as shown in the inset of Fig. 2(b).  
Crucially, the dependence of the elastic barrier on applied deformation is much stronger since both the harmonic spring constant and jump distance decrease with stress, and hence the elastic barrier is destroyed much rapidly with increasing stress or force compared to its local cage analog, as shown in the inset of Fig. 2(b). 
This means the theory predicts the degree of dynamic cooperativity of the alpha relaxation process decreases under deformation, i.e. the activated dynamics becomes more spatially local~\cite{ghosh2020a}.

In prior work and in the results shown in Figure 2 ~\cite{ghosh2020a,kobelev2005a}, $S(k)$ was assumed to be invariant to applied deformation. 
A schematic of the evolution of the dynamic free energy with increasing deformation (force) under this condition is shown in the main frame of Fig. 2b. 
The black curve corresponds to when the stress first destroys the barrier (where $r_L=r_B$), the ‘absolute yielding’ condition. 
The dynamically relaxed elastic shear modulus at the NMCT level is calculated from knowledge of the stress-dependent localization length using the expression ~\cite{ghosh2020a,kobelev2005a,naegele1998a},
\begin{equation}
\begin{split}
    G^\prime[\Sigma] = \frac{k_BT}{60\pi^2} \int_0^\infty ~dk \left[k^2\frac{d\ln{S({k})}}{dk} \right]^2 \exp{\left(-\frac{k^2r_L(\Sigma)^2}{3S(k)}\right)}\\
    \approx c\frac{\phi k_BT}{\sigma r_L(\Sigma)^2}
\end{split}
\tag{6}
\end{equation}
The last approximate equality is an analytic micro-rheology like relation previously derived under equilibrium conditions and verified to also hold well under stress~\cite{ghosh2020a}; $c$ is a known numerical constant. 
Eq (6) is directly relevant to real world mechanical measurements performed at finite frequencies only beyond the ideal NMCT “transition” ($r_L$ finite) in absence of ergodicity restoring activated events, and thus connects to the concept of “absolute yielding” mentioned above.

\section{Rheological Framework, the Isostructural Limit, and Beyond}
We first briefly review the adopted rheological framework in the context of the previously developed isostructural formulation of ECNLE theory under deformation~\cite{ghosh2020a}. 
We then motivate the need to consider deformation-induced changes of $S(k)$ to properly capture the overshoot physics within the same theoretical description.

Our starting point is a minimalist and physically transparent generalized Maxwell constitutive equation where the two key properties are the stress-dependent shear elastic modulus ($G^\prime$) and the structural or alpha relaxation time ($\tau_\alpha$) which we adopt as the characteristic mean stress relaxation time. 
For a constant shear rate $\dot{\gamma}$ the time (or strain)-dependent nonlinear stress relaxation function ($G$) obeys the first-order evolution equation~\cite{chen2011theory} $\displaystyle \frac{dG[\gamma,\Sigma(\gamma)]}{d\gamma} = - \frac{G[\gamma,\Sigma[\gamma]]}{\dot{\gamma}\tau_{\alpha}[\Sigma(\gamma)]}$, where $\gamma=\dot{\gamma}t$ is the accumulated strain and $\Sigma$ the stress. 
Formal integration through transients yields~\cite{chen2011theory},
\begin{equation}
    \Sigma(\gamma) = \int_0^{\gamma} d\gamma^\prime G^\prime[\Sigma(\gamma^\prime)] \exp{\left( -\int_{\gamma^{\prime}}^{\gamma} d\gamma^{\prime\prime} \frac{1}{\dot{\gamma}\tau_\alpha[\Sigma(\gamma^{\prime\prime})]} \right)} 
    \tag{7a}
\end{equation}
or equivalently in differential form,
\begin{equation}
    \frac{d\Sigma(\gamma)}{d\gamma}+\frac{\Sigma(\gamma)}{\dot{\gamma}\tau_\alpha[\Sigma(\gamma)]} = G^\prime[\Sigma(\gamma)]
    \tag{7b}
\end{equation}
From Eq.(7b), the steady state stress obeys the self-consistent equation, $\displaystyle \Sigma_{\text{ss}} = \dot{\gamma} \tau_\alpha[\Sigma_{\text{ss}}] G^\prime [\Sigma_{\text{ss}}]$.

We again note that, as also true of work of others (e.g., the original formulations of the soft glass rheology model~\cite{falk1998a}, the “isotropically-sheared” MCT approach~\cite{fuchs2002b}, NLE theory~\cite{kobelev2005a}), explicit tensorial anisotropy is ignored for tractability reasons.
This simplification is consistent with the description of colloid trajectories in terms of a scalar displacement dynamical variable in NLE theory, and the adoption of an isotropic structure factor, $S(k)$, to construct the dynamic free energy. Of course, neglecting explicit anisotropy is an approximation, but the practical question is how large are the consequences on the cage scale for the scalar observables of interest. 
In this regard, simulation studies of the rheology of glassy fluids~\cite{miyazaki2004a,yamamoto1998a} found that even in the strong shear thinning regime, for cage scale wavevectors, $k$, that define structural relaxation, which is strongly coupled with stress relaxation, there is very little dynamical anisotropy of the measured alpha time, mean square displacement, and incoherent dynamic structure factor, $F_s (k,t)$.
This supports, or at a minimum can rationalize, why using an effectively isotropic $S(k)$ can successfully capture many aspects of glassy colloid rheology.

Now, assuming literally no change of $S(k)$ under deformation~\cite{ghosh2020a}, ECNLE theory predicts the required elastic modulus and alpha time in Eq.(7) as a function of stress. 
Deformation enters $F_{\text{dyn}} (r)$ solely in a mechanical work like manner (Fig. 2) and monotonically reduces $\tau_\alpha$ and $G^\prime$ with increasing stress as calculated using the Kramers’ formula for barrier crossing~\cite{mirigian2014a,ghosh2020a,schweizer2005a} and the Green-Kubo-like elastic shear modulus expression~\cite{kobelev2005a,naegele1998a} discussed above. 

The core hypotheses in our present new work are: (1) within the generalized Maxwell constitutive equation approach, structure changes under deformation are necessary to obtain an overshoot in the stress- strain response, and (2) for a steady state stress to exist, structural changes must eventually stop at sufficiently high strain or time.
We justify these hypotheses by analyzing the generalized Maxwell constitutive equation. 
First, since $\displaystyle \frac{d\Sigma(\gamma)}{d\gamma} =  G^\prime[\Sigma(\gamma)] - \frac{\Sigma(\gamma)}{\dot{\gamma}\tau_\alpha[\Sigma(\gamma)]}$, for the invariant structure model if $\displaystyle \frac{d\Sigma(\gamma)}{d\gamma} = 0$ then strain becomes an irrelevant variable and one has $\displaystyle G^\prime[\Sigma] = \frac{\Sigma}{\dot{\gamma}\tau_\alpha[\Sigma]}$.
With increasing stress at a fixed shear rate, the LHS side of this equation monotonically decreases, while the RHS monotonically increases (the alpha time is a monotonically decreasing function of applied force), and thus only one crossing of curves (the mathematical solution) is possible which defines the steady state.
This proves that within our formulation it is impossible to have an overshoot if structure does not change (corresponding to an ideal plastic response), consistent with point (1). 

For a changing structure under deformation model, using again $\displaystyle \frac{d\Sigma(\gamma)}{d\gamma} =  G^\prime[\Sigma(\gamma)] - \frac{\Sigma(\gamma)}{\dot{\gamma}\tau_\alpha[\Sigma(\gamma)]}$, one finds the second derivative of stress with respect to strain is,
\begin{equation*}
    \begin{split}
        \frac{d^2\Sigma(\gamma)}{d\gamma^2} = \left( \frac{\partial G^\prime[\Sigma(\gamma)]}{\partial \gamma} + \frac{\Sigma(\gamma)}{\dot{\gamma}\tau_\alpha[\Sigma(\gamma)]^2}\frac{\partial \tau_\alpha[\Sigma(\gamma)]}{\partial \gamma}\right) 
        + \frac{d\Sigma(\gamma)}{d\gamma}\times \\ \left[\frac{dG^\prime[\Sigma(\gamma)]}{d\Sigma} -  \frac{1}{\dot{\gamma}\tau_\alpha[\Sigma(\gamma)]}+\frac{\Sigma(\gamma)}{\dot{\gamma}\tau_\alpha[\Sigma(\gamma)]^2}\frac{d \tau_\alpha[\Sigma(\gamma)]}{d \Sigma}\right]
    \end{split}
    \tag{8}
\end{equation*}
The partial derivatives come from purely structural changes with deformation. For both the steady state and at the stress overshoot one has $\displaystyle \frac{d\Sigma(\gamma)}{d\gamma} = 0$, and thus only the partial derivative terms survive. 
Thus, for both the steady state and overshoot maximum one can write,
\begin{equation}
    \frac{d^2\Sigma(\gamma)}{d\gamma^2}\bigg|_{\gamma=\gamma_{\text{pk}},\gamma_{\text{ss}}} = \left( \frac{\partial G^\prime[\Sigma(\gamma)]}{\partial \gamma} + \frac{\Sigma(\gamma)}{\dot{\gamma}\tau_\alpha[\Sigma(\gamma)]^2}\frac{\partial \tau_\alpha[\Sigma(\gamma)]}{\partial \gamma}\right)
    \tag{9}
\end{equation}
An additional criterion for the overshoot is $\displaystyle \frac{d^2\Sigma(\gamma)}{d\gamma^2}\bigg|_{\gamma=\gamma_{\text{pk}}}< 0$. 
In general, $G^\prime[\Sigma(\gamma)]$ decreases with stress and strain-induced softening of structure, so its partial derivative in Eq. 9 is always negative. 
The same is true for $\displaystyle \frac{\partial \tau_\alpha[\Sigma(\gamma)]}{\partial \gamma}$ (the numerical prefactor is positive), and hence $\displaystyle \frac{d^2\Sigma(\gamma)}{d\gamma^2}<0$. 
This proves that structural changes are necessary for the partial derivatives to be nonzero, a necessary condition for a negative second derivative. This completes our discussion of point (1) above.

A steady state requires $\displaystyle \frac{\partial G^\prime[\Sigma(\gamma)]}{\partial \gamma} = 0$ and $\displaystyle \frac{\partial \tau_\alpha[\Sigma(\gamma)]}{\partial \gamma}=0$ for $\displaystyle \frac{d^2\Sigma(\gamma)}{d\gamma^2}=0$. 
This is impossible if deformation-induced changes in $S(k)$ do not eventually vanish at long time (large strains), which is physically obvious.  
To proceed, we first assume an overshoot stress maximum exists and occurs at $\gamma=\gamma_m$ where all the above-mentioned criteria are satisfied. 
We explore what happens if one postulates that structure does not change (achieves its nonequilibrium steady state) for $\displaystyle \lim_{\epsilon\to 0^{+}} \frac{d^2\Sigma(\gamma)}{d\gamma^2}\bigg|_{\gamma=\gamma_m+\epsilon} = 0$ as the partial derivatives in Eq. 9 are zero, and this is the criterion for a steady state. 
Thus, no overshoot in the stress strain curve is possible, and to have an overshoot structural changes must continue beyond the overshoot. 
We believe this to be a physically intuitive condition since our view is that the overshoot qualitatively indicates a crossover from an elastic to viscous like mechanical response. 
Simulation findings [16-18] are also consistent with this perspective since they find structural changes do not saturate at the stress overshoot, but rather require strains modestly beyond it. 
This completes our discussion of point (2).

\section{Structural Changes with Deformation}

In Brownian suspensions not close to a shear jamming or thickening regime, the cage scale packing correlations, and hence kinetic constraints, are generally believed to on average weaken under shear deformation~\cite{larson1999structure,koumakis2016a,koumakis2012a,fuchs2002a,fuchs2005a}. 
As mentioned above, simulations find that the shear rate dependences of the local structure correlations, the anisotropy of particle diffusion (after subtracting kinetic advection effects) and the incoherent dynamic structure factor that quantifies cage scale structural relaxation, are very small even under strong deformation~\cite{miyazaki2004a,yamamoto1998a}. 

Here we construct a simple and physically transparent treatment of this key structural softening effect in real space which enters ECNLE theory and the nonlinear Maxwell model via an isotropic but strain-dependent $S(k,\gamma)$ roughly in the spirit of an elastic-like analog of the shear advection~\cite{fuchs2002a,amann2013a,fuchs2002a,fuchs2005a} idea. 
Within this simplified approach that ignores explicit anisotropic changes of structure, and as discussed in textbooks, one expects that under deformation the average separation between a pair of particles (here hard spheres, diameter $\sigma$) initially increase affinely with strain~\cite{larson1999structure}. 
The radial distribution function (and thus $S(k)$ and the direct correlation function, $C(k)$) becomes strain-dependent which is modeled in an effectively isotropic manner as: $\displaystyle g_{\gamma}(r\ge \sigma)=g_0\left(r\sqrt{1+\gamma^2/3}\right)$, where $g_0 (r)$ is the equilibrium pair correlation function and $g_\gamma(r<\sigma)=0$.
Strain reduces the contact value to $\displaystyle g_0\left(\sigma\sqrt{1+\gamma^2/3}\right)$, and more generally weakens the structural pair correlations. 
The dynamic theory requires as input the structure factor, $\displaystyle S(k,\gamma)=1+\rho h(k,\gamma)$, and hence one needs,
\begin{equation*}
    \begin{split}
        h(k,\gamma) = \int d\vec{r} \exp{(-i\vec{k}\cdot\vec{r})}h\left(\vec{r}\sqrt{1+\gamma^2/3}\right)\\
        = \int \frac{d\vec{R}}{(1+\gamma^2/3)^{3/2}}\exp{\left(-i\frac{\vec{k}}{\sqrt{1+\gamma^2/3}}\cdot \vec{R}\right)}\\
        = \frac{1}{(1+\gamma^2/3)^{3/2}}h_0\left(\frac{{k}}{\sqrt{1+\gamma^2/3}}\right)
    \end{split}
    \tag{10}
\end{equation*}
This gives the final form of the deformation-dependent structure factor $S(k,\gamma)$ we employ as,
\begin{equation*}
    \begin{split}
        S(k,\gamma)= 1+\frac{\rho}{(1+\gamma^2/3)^{3/2}}h_0\left(\frac{{k}}{\sqrt{1+\gamma^2/3}}\right)\\
        = S_0(k) + \rho\left(\frac{h_0\left(\frac{k}{\sqrt{1+\gamma^2/3}}\right)}{(1+\gamma^2/3)^{3/2}}-h_0(k)\right)\\
        \approx S_0(k) +\rho h_0(k)\left((1+\gamma^2/3)^{-3/2}-1\right)
    \end{split}
    \tag{11}
\end{equation*}
The last approximate relation is written only to allow one to more easily see how strain modifies the wavevector dependence (it is mathematically exact within this model for $k=0$). 
In all numerical calculations we employ the full relation. 
The $k\to 0$ limit quantifies the density fluctuation amplitude (or thermodynamic “dimensionless compressibility”) and is given as,
\begin{equation}
    S(k\to 0, \gamma) = S_0(0) +\rho h_0(0)\left((1+\gamma^2/3)^{-3/2}-1\right)
    \tag{12}
\end{equation}
Since generically both $\displaystyle (1+\gamma^2/3)^{-3/2}-1<0$, and $\displaystyle \rho h(0)<0$, one has $\displaystyle S(k\to 0, \gamma) > S_0(0)$.
Thus, strain weakens local packing correlations on the cage scale, and also enhances the amplitude of long wavelength density fluctuations in qualitative analogy with a system being at an effectively lower packing fraction. 
The above model connects in spirit with prior NLE theory rheology work for polymer glasses that adopted a coarse-grained chain model with no local liquid like structure~\cite{chen2011theory}. 
In that approach, an increase of $S(k=0)$ under deformation [called a “mechanical disordering or rejuvenation effect”] below $T_g$~\cite{chen2011theory} results in the prediction of an overshoot in the transient response. 
Here for colloidal systems, aging is not considered, and we allow deformation to change packing structure on all length scales, with changes on the local cage scale most important.  

\begin{figure*}
    \centering
    \includegraphics[width=0.95\textwidth]{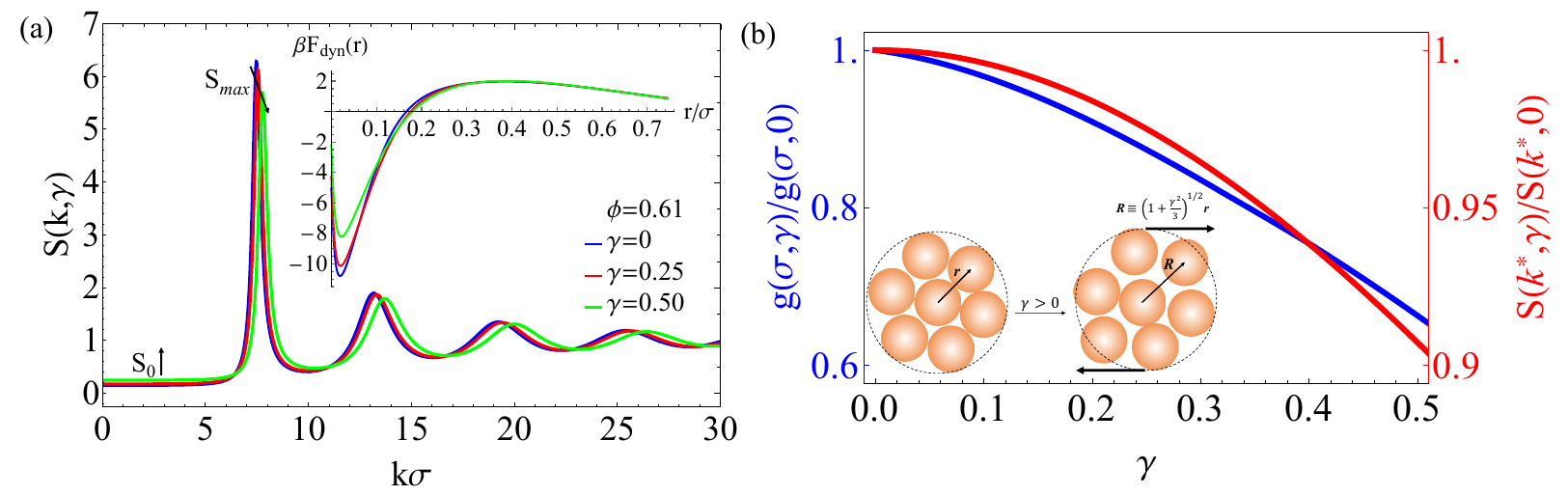}
    \caption{(a) Static structure factor at a fixed volume fraction for various strain values. With increasing strain ($\gamma$), an increase in $S(k\to 0)=S_0$ and a decrease in $S(k=k^\star)=S_{\text{max}}$ is observed. Implications of the changing structure model on the dynamic free energy is shown in the inset. (b) Quantification of the softening of two local structural metrics: normalized contact value (left axis) and normalized cage coherence peak (right axis), for a packing fraction of 0.61. Inset shows a schematic of the cage ‘distortion’ model that motivates the choice of renormalized average distance between particles as a function of strain and the isotropic scalar separation of caged particles. }
    \label{fig:sk_gr}
\end{figure*}
Figure 3a shows the strain dependent structure factor for a fixed packing fraction of $0.61$. Per Eq. 12, $S(k=0,\gamma)$ is an increasing function of strain $\gamma$, while the cage coherence peak decreases indicating cage weakening. 
Figure 3b shows the normalized contact value of the radial distribution function monotonically decreases with strain. 
A normalized “cage order parameter” is defined as the ratio $S(k^\star,\gamma)/S(k^\star,0)$ where $k^\star$ is the wavevector at the first maximum, and it also decreases with strain (right legend).
If packing fraction is reduced, nearly identical curves are found in this normalized framework (not shown), and the results in Fig. 3b are thus generic.

The deformation-dependent $S(k)$ enters the calculation of the dynamic free energy, from which all dynamical properties relevant to our rheological predictions follow using ECNLE theory. 
The new feature of the changing structure model is deformation modifies both the force vertex (kinetic constraints, second term in Eq. 2) of the dynamic free energy and contributes the external force or stress dependent mechanical work term in the nonequilibrium dynamic free energy of Eq. 5. 
Hence all theoretical quantities relevant to dynamics and rheology depend directly on $\Sigma$ and $\gamma$.

\section{Effect of Deformation on the Relaxation Time and Elastic Modulus and Onset of the Structural Steady State}
Our modeling of structural softening weakens all the localizing features of the dynamic free energy, resulting in strain-dependent decreases of $\tau_\alpha$ and $G^\prime$, in addition to the direct effect of stress via the mechanical work term in the nonequilibrium dynamic free energy. 
Representative results using Eqs. 4 and 6 for the relaxation time and elastic modulus are plotted as a function of stress and strain in Fig. 4 based on treating them as independent variables. 
For the deformation invariant structure model, the elastic modulus decreases relatively slowly (the $\gamma=0$ result is the thick blue curve in Fig. 4(b)).
The new rheologically consistent approach (i.e., using the predicted stress-strain curve) with structural changes corresponds to traversing a specific path in $(G^\prime,\Sigma,\gamma)$ space as shown by the dashed curve for a dimensionless shear rate of $0.01$. The same illustrative path is shown for the alpha relaxation time in panel (a).

\begin{figure*}
    \centering
    \includegraphics[width=0.95\textwidth]{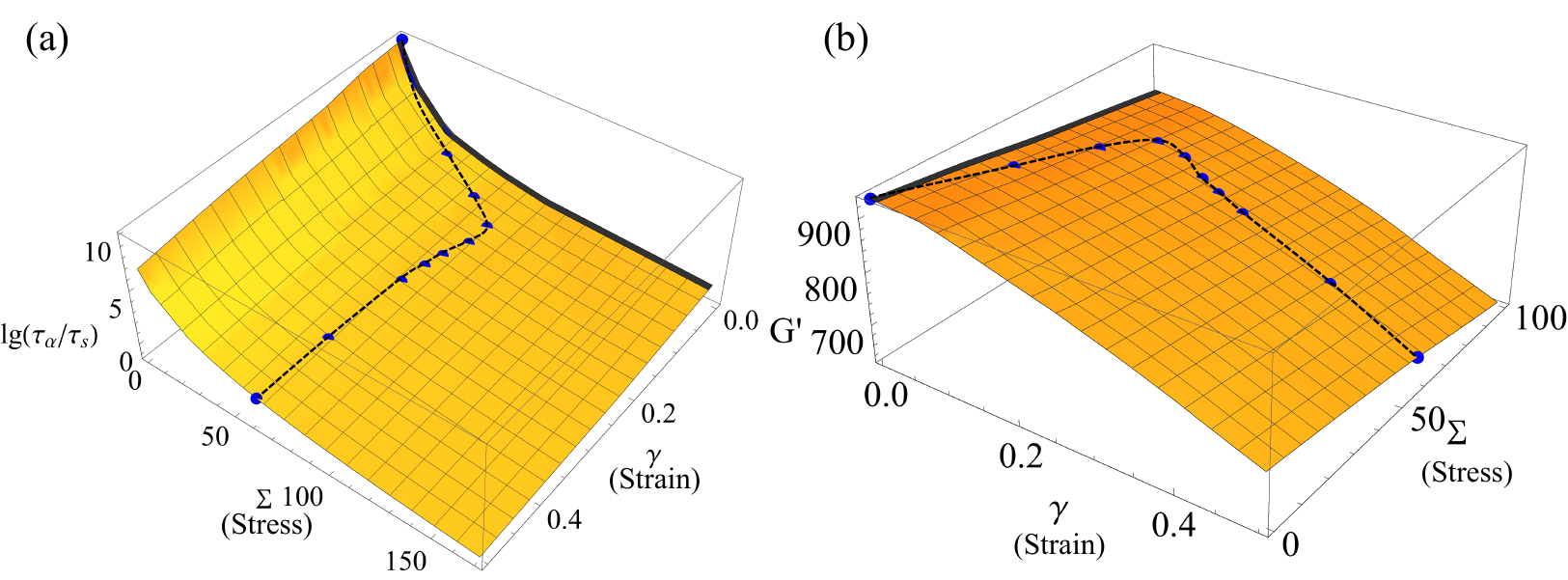}
    \caption{Stress and strain dependent (a) alpha or structural relaxation time, and (b) elastic shear modulus for a packing fraction of 0.61. A rheologically consistent treatment follows a specific path determined from the theoretical constitutive equation as shown above for a dimensionless shear rate of 0.01. Thick solid curves at $\gamma=0$ represent how the relaxation time and elastic modulus change as a function of stress for a deformation-invariant structure model.}
    \label{fig:fig_3d}
\end{figure*}

While a complete microscopic theory of the full stress strain response also requires a self-consistent equation for the evolution of the static structure factor under deformation, we proceed here by formulating a simpler and far more computationally tractable approach to quantify when deformation no longer changes structural correlations. 

\begin{figure*}
    \centering
    \includegraphics[width=0.95\textwidth]{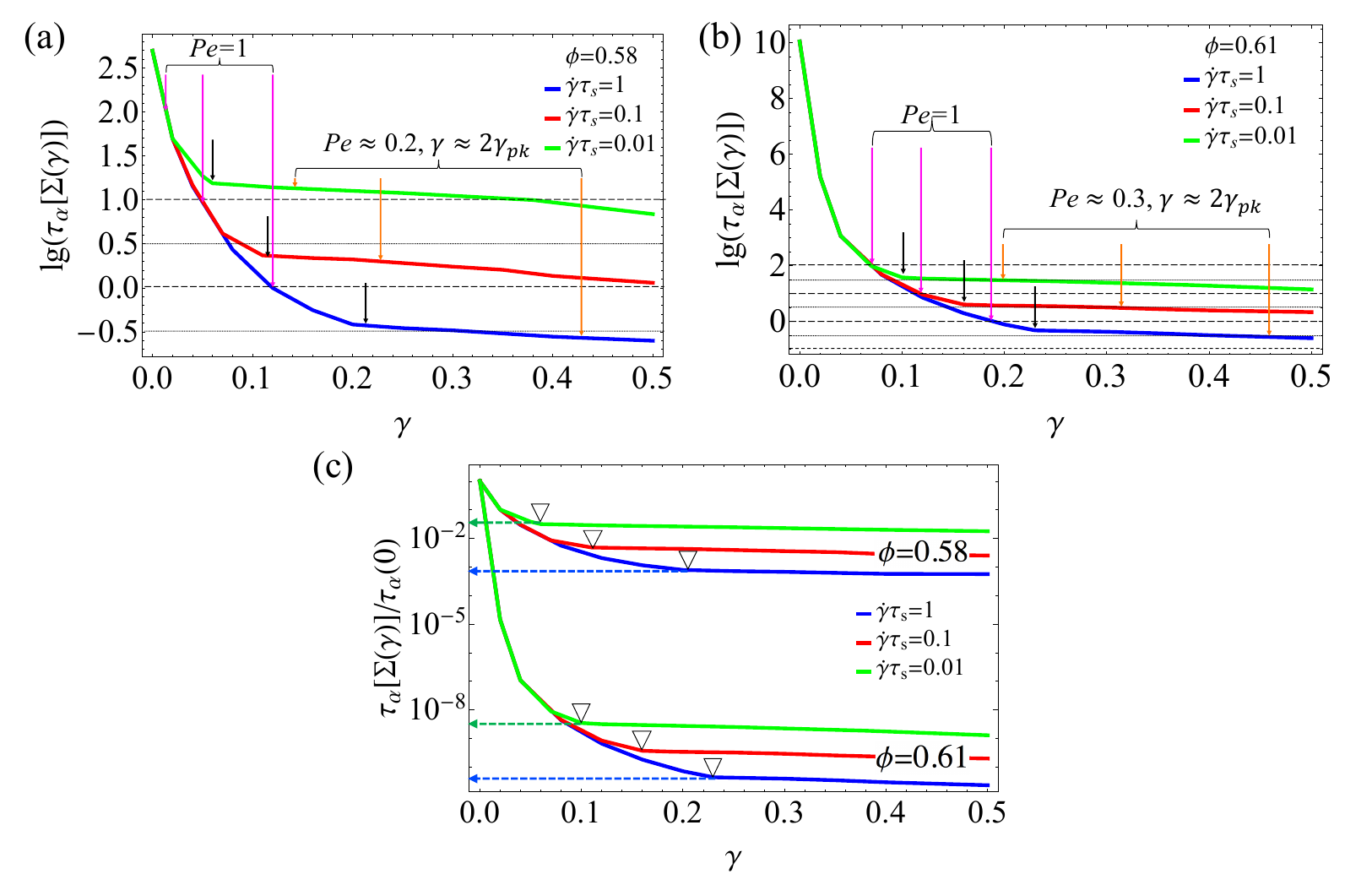}
    \caption{Evolution of the alpha relaxation time as a function of strain for a packing fraction of (a) 0.58 and (b) 0.61 at several dimensionless shear rates relevant to experimental studies~\cite{koumakis2012a,koumakis2016a,laurati2012a}. The black arrows indicate the overshoot strain ($\gamma_{\text{pk}}$). Magenta and orange arrows indicate two different strains where $\text{Pe}=\dot{\gamma}\tau_\alpha[\Sigma(\gamma)]=1$ and $\gamma=2\gamma_{\text{pk}}$ respectively. Panel (c) shows the normalized alpha relaxation times for two different packing fractions of 0.58 and 0.61. The downward triangles indicate overshoot location and horizontal arrows denote orders of magnitude faster dynamics with respect to the “equilibrium” alpha relaxation ($\gamma=0$).}
    \label{fig:tau_alpha}
\end{figure*}
Figure 5a and 5b plots the logarithm of the alpha time as a function of strain for packing fractions of $0.58$ and $0.61$ and several shear rates. 
The overshoot strains ($\gamma_{\text{pk}}(\dot{\gamma})$) are indicated by vertical black arrows. 
Magenta arrows show when the renormalized Pe equals unity in both plots, defined as at $\gamma=\gamma_{\text{Pe}=1}$. 
We note that setting $S(k,\gamma\ge \gamma_{\text{Pe}=1})=S(k, \gamma_{\text{Pe}=1})$ always destroys the overshoot, hence $\gamma_{\text{Pe}=1}>\gamma_{\text{pk}}$ is required. 
This inequality is consistent with the experimental and simulation findings for hard sphere suspensions~\cite{koumakis2012a,koumakis2016a,laurati2012a} that at the overshoot $\text{Pe}(\gamma_{\text{pk}}(\dot{\gamma}))<1$. 
These same studies~\cite{koumakis2012a,koumakis2016a,laurati2012a} find the structural steady state is effectively achieved at strains modestly beyond the overshoot of $b\gamma_{\text{pk}}$, i.e., $S(k,\gamma\ge b\gamma_{\text{pk}})=S(k,b\gamma_{\text{pk}})$, where $b\sim 2-3$~\cite{jadhao2017probing,pham2006yielding}. 
These considerations are what motivated our modeling of structure change under deformation discussed in the previous section.

With the above motivation, exploratory calculations with the new theory including $S(k,\gamma)$ have been performed over a wide range of shear rates and packing fractions. 
We find that a sensible criterion for the emergence of a steady state structure is $\text{Pe}(2\gamma_{\text{pk}}(\dot{\gamma}|))=a(\phi)\approx 0.15-0.30$ independent of shear rate to zeroth order, and very weakly dependent on packing fraction for $\phi\in (0.55,0.61)$ which is a range that corresponds to the quiescent alpha time varying by ~9 decades~\cite{mirigian2014a,ghosh2020a}. 
We thus take as our core hypothesis that structure becomes effectively deformation-independent at a single fixed value of renormalized $\text{Pe}$: $S(k,\gamma\ge \gamma_c)=S(k,\gamma_c)$ where $\gamma_c$ is determined from the criterion $\text{Pe}(\gamma_c) = \dot{\gamma}\tau_\alpha[\Sigma(\gamma_c)]=C$, with $C$ a universal constant for a given material system. 
We have verified that different choices of $C\in (0.1,0.3)$ only very weakly modify the results discussed below, which are shown using $C=0.25$\footnote{We begin to see deviations beyond $A=0.3$ where the magnitude of the overshoot (steady state stress) starts to decrease (increase). Eventually for $A\ge 0.5$ the overshoot is almost destroyed.}. 
This fixed $\text{Pe}$ criterion introduces just one adjustable number of clear physical meaning. We find the above criterion is equivalent to $\gamma_c/\gamma_{\text{pk}}\sim 1.8-2.3$ for all packing fractions and shear rates studied. 
Nontrivially, the above perspective agrees well with experiments and simulations on ultra-dense colloidal suspensions, and also prior simulations~\cite{hoy2010a}, experiments~\cite{bennin2019a,bending2014a,lee2010a,pamvouxoglou2021a}, and theoretical~\cite{chen2011theory} analysis of glassy polymers which find $Pe(\gamma_c )\sim 0.2\pm 0.05$. 

\section{Theory Predictions}
\subsection{Alpha Relaxation}
ECNLE theory predicts~\cite{mirigian2014a} the quiescent alpha relaxation slows down by ~6 decades as the packing fraction grows from 0.58 to 0.61. 
The theory also predicts massive speed up of the alpha time under deformation~\cite{ghosh2020a}. Since structural correlations soften the caging constraints, the degree of speed-up in dynamics must be stronger than predicted based on using the invariant structure model~\cite{ghosh2020a}. 
Figure 5c presents the normalized relaxation times for two packing fractions and several strain rates. 
At $\gamma\tau_s=1$, alpha relaxation time at the overshoot speeds up by $\sim 2$ or $\sim 8$ orders of magnitude for packing fractions of 0.58 or 0.61, respectively. 
At a lower strain rate of 0.01, the relaxation time speeds up by $\sim 3$ to $\sim 9.5$ decades for the same packing fractions. 
As shown in Fig. 5b, for both the packing fractions studied and several strain rates, the relaxation time at the overshoot point and in the steady state ($\gamma_{\text{ss}}\sim [2-3]\gamma_{\text{pk}}$) are almost exactly the same. 

\begin{figure*}
    \centering
    \includegraphics[width=0.95\textwidth]{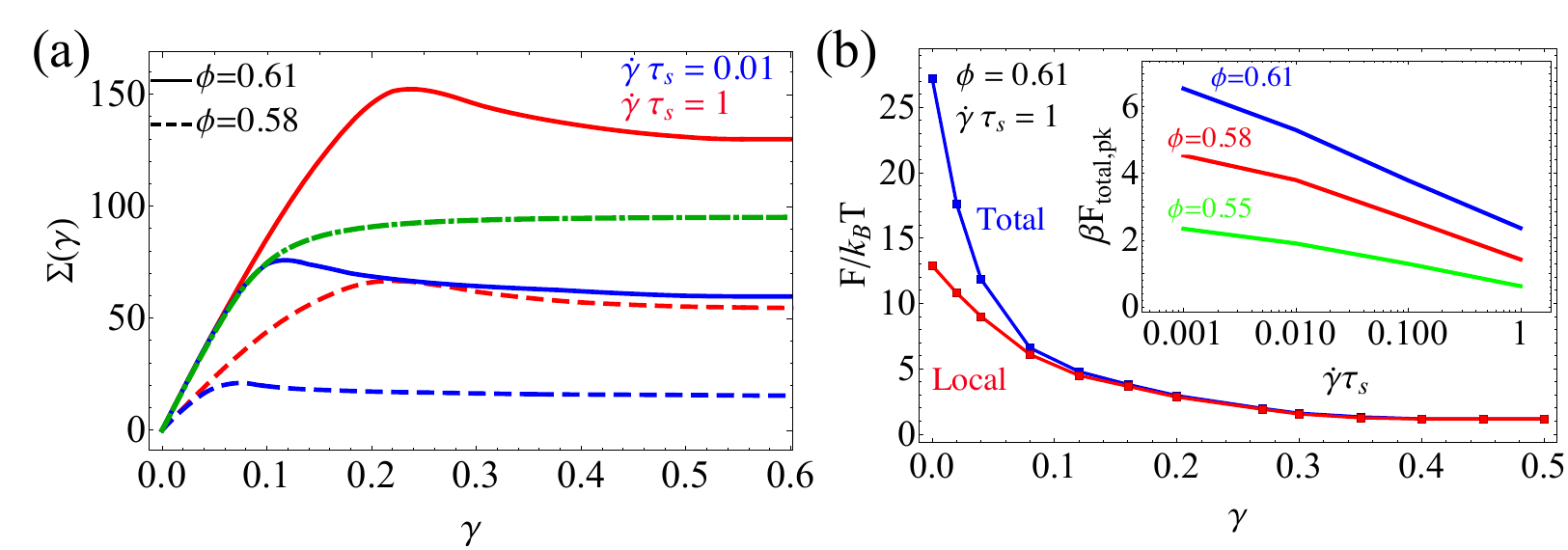}
    \caption{: (a) Theoretical stress (unit of $k_BT/\sigma^3$) vs strain curve for packing fractions 0.58 (dashed) and 0.61 (solid) and two reduced shear rates~\cite{schweizer2003a}. The green dash-dotted curve is for $\phi=0.61$ and $\dot{\gamma}\tau_s=0.01$ using the deformation-invariant structure model and does not exhibit an overshoot (compare to the blue solid curve). (b) Total and local cage barriers in thermal energy units as a function of strain. At the overshoot ($\gamma_{\text{pk}}\sim 0.22$) the total barrier is predominantly the caging contribution. Inset: total barrier at the overshoot as a function of shear rate for different packing fractions.}
    \label{fig:overshoot}
\end{figure*}

\subsection{Overshoot Predictions: Full Rheological Response and Role of Activation}
Figure 6a shows representative stress-strain curves for packing fractions and shear rates~\footnote{Experiments and simulations sometimes employ a “bare” Peclet number $\text{Pe}_0=\dot{\gamma}\tau_0$, where $\tau_0$ is the dilute suspension Brownian time. Many other workers adopt a known short time process time scale (includes local hydrodynamics or binary collisions) that obeys $\tau_0 < \tau_s \ll \tau_\alpha$. All our theoretical analysis uses this short time Peclet number, $\text{Pe}_s=\dot{\gamma} \tau_s$, to display our results.} typical of ultra-dense colloidal suspension experiments~\cite{koumakis2016a,koumakis2012a,laurati2012a}.
All exhibit stress overshoots, which reflects a competition between the “direct” effect of stress via the mechanical work term in the nonequilibrium dynamic free energy and the softening of caging constraints with strain as encoded in $S(k,\gamma)$ and $C(k,\gamma)$. 
Specifically, although stress and strain both accelerate relaxation and soften the elastic modulus, they do so in different ways, and these effects are self-consistently coupled via the nonlocal (strain history integration) and nonlinear form of Eq.(7). 

Fig.6a shows that in the low-strain elastic regime, $\Sigma$ grows with strain in a shear-rate independent manner. 
A representative example of our general finding that the deformation-invariant structure model ($S(k,\gamma)=S(k,\gamma=0)$) predicts no overshoot is also presented, consistent with our analytic arguments in section III. 
The steady state stress is smaller in the presence of structural softening, as expected. 

To understand the importance of activated dynamics in general, and specifically the role of collective elasticity on the activation barrier, we plot the total and local cage barriers as a function of strain in Fig. 6b, and the total barrier at the overshoot peak as a function of shear rate. 
The total barrier is not negligible at the overshoot, and is larger at higher packing fractions and lower shear rates. 
This existence of a nonzero barrier is consistent with experimental and simulation observations that intermittent trajectories persist under strong deformation~\cite{besseling2007three,hoy2010a}. The main frame of Fig.6b also shows the total barrier is dominated by its local cage component at and beyond the overshoot. 
This is a consequence of the predicted ease of destruction of the elastic barrier under deformation~\cite{ghosh2020a}. 

At very large shear rates (typically beyond experimental study), the inset of Fig.5b suggests a type of “granular” or “non-Brownian” yielding may apply in the sense that the barrier at the overshoot (dynamic yield point) is asymptotically approaching zero, and hence no thermal activation is required (per an ideal MCT scenario). 
However, as established below, the strong rate and packing fraction dependences of the barrier do critically impact our predictions for how key rheological signatures evolve with these control parameters. 

\subsection{Correlation Between Peak Curvature and Overshoot Magnitude}
\begin{figure}
    \centering
    \includegraphics[scale=0.85]{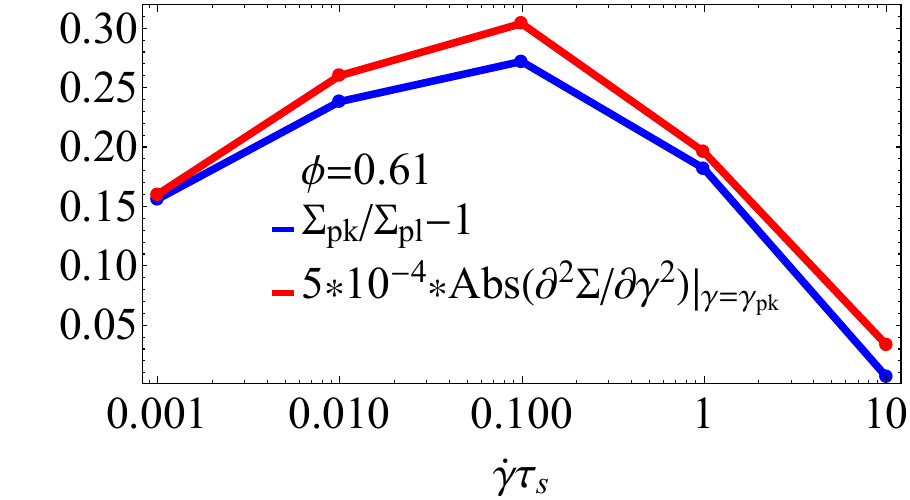}
    \caption{Absolute magnitude of the second derivative of the stress-strain curve at the overshoot and the overshoot amplitude as a function of shear rate for a packing fraction of 0.61.}
    \label{fig:corr}
\end{figure}

Our intuition based on the analytic analysis presented in section III is that the absolute magnitude of the second derivative of the stress-strain curve at the overshoot and the amplitude of overshoot are highly correlated. 
To test this idea numerically, we plot in Fig. 7 the rescaled absolute value of the 2nd derivative and overshoot amplitude as a function of shear rate for fixed packing of 0.61. One sees the two curves track each other extremely well, indicating a strong casual correlation between them as, $\displaystyle \bigg|\frac{\partial^2 \Sigma}{\partial \gamma^2}\bigg|_{\gamma=\gamma_{\text{pk}}}\approx a+b\left(\Sigma_{\text{pk}}/\Sigma_{\text{pl}}-1\right)$ where $a$ and $b$ are constants. 
This example for a packing fraction of $0.61$ is not special, and the same level of correspondence is found for all the other high packing fractions studied.

\subsection{Steady State Analysis}
\begin{figure}
    \centering
    \includegraphics[scale=0.27]{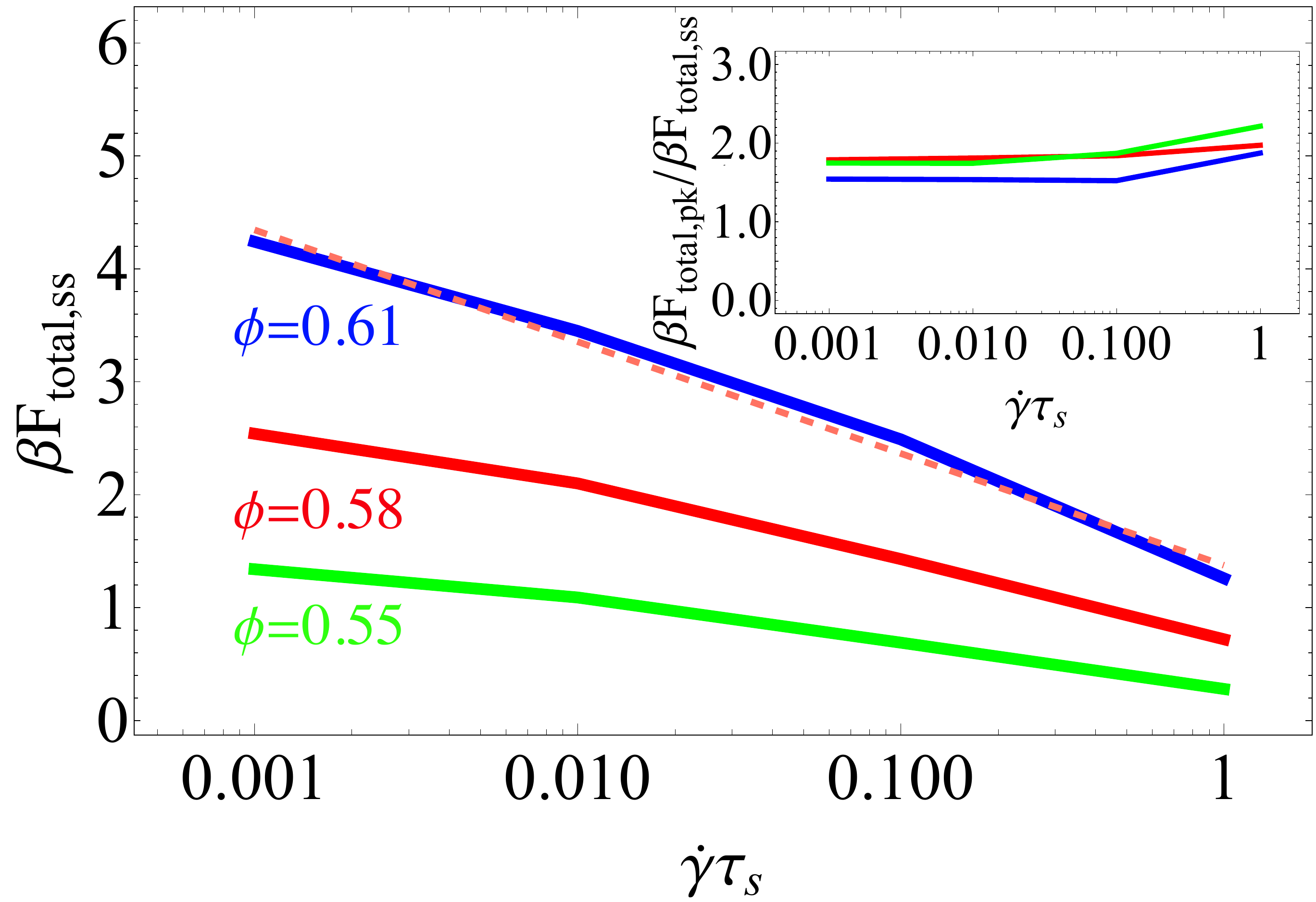}
    \caption{Total activation barrier (thermal energy units) in the steady state as a function of dimensionless shear rate for a wide packing fraction range. The dotted orange dashed straight line shows a $\beta F_{\text{total, ss}}\sim -\log{(\dot{\gamma}\tau_s)}$ fit of the numerical calculations. Inset: ratio of the total barrier at the overshoot to that in the steady state as a function of dimensionless shear rate. Colors have the same meaning as in the main panel.}
    \label{fig:f_totvsgammadot}
\end{figure}
Results for the total activation barrier as a function of shear rate and packing fraction in the steady state are shown in Fig. 8. 
This total barrier is dominated by the local caging contribution since the collective elastic effects become small beginning at low strains or stresses, as previously discussed~\cite{ghosh2020a}. 
However, a non-vanishing total barrier is predicted even in the steady state, thereby establishing that some level of activated dynamics remains present. 
The steady state total barrier roughly follows a logarithmic decrease with dimensionless shear rate, $\displaystyle \beta F_{\text{total,ss}}\sim -\ln{(\dot{\gamma}\tau_s)}$, as illustrated by the dashed blue line in Fig. 8 for a packing fraction of $0.61$. 
Interestingly, the inset of Fig. 8 shows that the ratio of total barrier at the stress overshoot to that in the steady state is predicted be roughly constant $\sim 1.5-2$. 
This represents a non-trivial connection of activated physics in the transient and steady state regimes. For dimensionless shear rates $>10$ (not shown) the steady state barriers become very small. 

Finally, we consider how deformation-induced changes of structure modifies steady state properties compared to our prior isostructural theory results~\cite{ghosh2020a} and experiment. 
Fig. 9 shows representative calculations of the flow curve (steady state stress vs shear rate) and shear thinning of the relaxation time. 
The prior prediction that the onset of shear thinning begins at remarkably low Pe values is robust to including deformation-induced structural changes, and agrees with simulations and experiments as previously discussed~\cite{ghosh2020a}. 
Only modest increases of the apparent thinning exponent are predicted when structural changes are included. 
The flow curves are also only very weakly affected by deformation-induced structural changes. Moreover, note that the empirical Herschel-Buckley (HB) form observed experimentally, $\Sigma_{\text{ss}}=\Sigma_{\text{ss}}^{\text{HB}}+c\dot{\gamma}^n$, is generically predicted by the microscopic theory, not assumed as in most phenomenological models. 
This is significant since the HB form is generally empirically adopted with adjustable parameters in the analysis of experimental and simulation data.  
Here, the HB form emerges in a predictive way from our approach based on analyzing our numerical results over a practical finite range of shear rates as done in experiment and simulation. 
Moreover, the idea there is a static yield stress that is nonzero in the literal zero shear rate limit is not true in our theory where there is always finite, albeit very slow, ergodicity restoring structural relaxation in the absence of a literal jamming point. 

\begin{figure}
    \centering
    \includegraphics[scale=0.85]{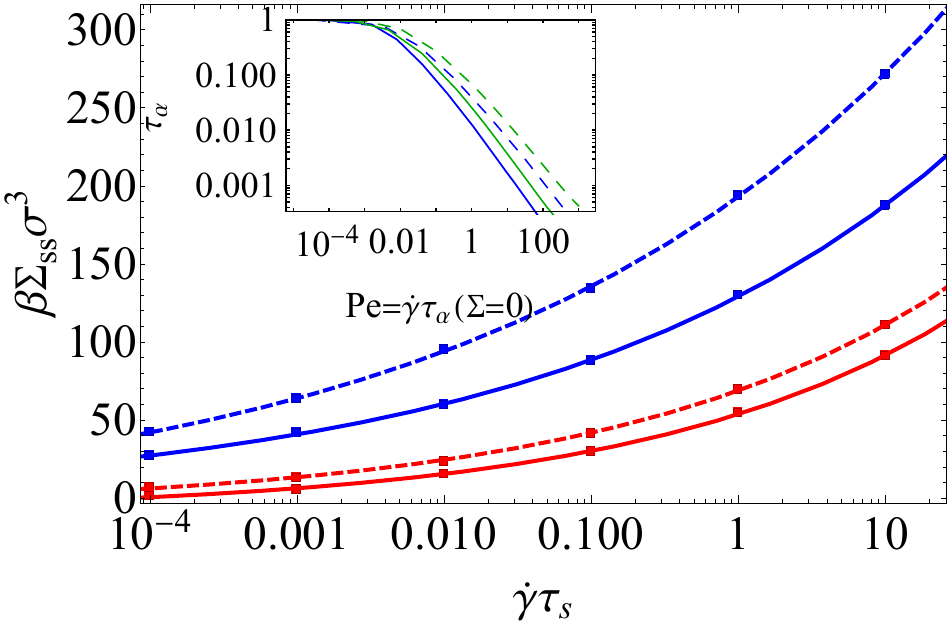}
    \caption{Non-dimensionalized steady state stress as a function of dimensionless shear rate (points) for packing fractions 0.61 (upper blue points) and 0.58 (lower red points); results are shown using the changing (lower stress points) and invariant with deformation structure models. The corresponding smooth (solid, dashed) curves are HB model fits for packing fractions of 0.61 (n=0.15 and 0.17) and 0.58 (n=0.20 and 0.21), where the HB exponents, n, are slightly smaller for the invariant structure model. Inset: steady state alpha time normalized to its quiescent value as a function of renormalized Peclet number for the invariant (dashed) and changing (solid) structure models for packing fractions of 0.61 (blue) and 0.58 (green).}
    \label{fig:hb}
\end{figure}

Finally, we note that additional analysis reveals a nearly universal value of roughly 10 for the ratio of local cage to collective elastic barriers at the overshoot and in the steady-state, and hence the total barrier is dominated by the local caging contribution. 
The relevant plot is not shown since the absolute value of the collective elastic barrier at these rheological state points is negligible ($F_e\ll k_BT$), consistent with the results in Figs. 2b, 6b and 8.

\section{Comparison with Experiments and Simulations }
\subsection{Stress Overshoot}
Figure 10 presents predictions (not fits) for the packing fraction and shear rate dependences of the stress overshoot amplitude, $r_O=\Sigma_{\text{pk}}/\Sigma_{\text{ss}}-1$ , and its corresponding strain, $\gamma_{\text{pk}}$; experimental results for hard sphere colloidal suspensions from refs~\cite{chen2011theory,koumakis2016a} are shown as discrete points. 
The dependence of $\gamma_{\text{pk}}$  on shear rate for different packing fractions is presented in Fig.10a. 
Experiments find a factor of $\sim 2.5$ increase (from $\sim 8$ to 20\%) for a $\sim$1-1.5 decade increase of shear rate over the range $\phi=0.54-0.6$, simulations find qualitatively similar results with $\gamma_{\text{pk}}\sim $ ~15-40\% for a 2 decade increase of shear rate~\cite{chen2011theory,koumakis2016a}, and the theory predicts a factor of $\sim 2-4$ growth of $\gamma_{\text{pk}}$ depending on packing fraction. The theoretical trend of smaller changes for higher packing fractions is consistent with simulation~\cite{chen2011theory,koumakis2016a}. 
We note that for $\phi=0.61$, the peak strain follows an apparent power law $\gamma_{\text{pk}}\sim \dot{\gamma}^{0.16}$ for all the relevant strain rates spanning 4 decades (black dotted line in Fig. 10 (a)), with a small effective exponent quite similar to our prediction for the HB exponent in the previous section. 
The same qualitative similarity remains true for a lower packing fraction of 0.58, where the exponent is $\sim 0.22$ for $\dot{\gamma}\tau_s\in(0.001,1)$ (not shown). 

\begin{figure*}
    \centering
    \includegraphics{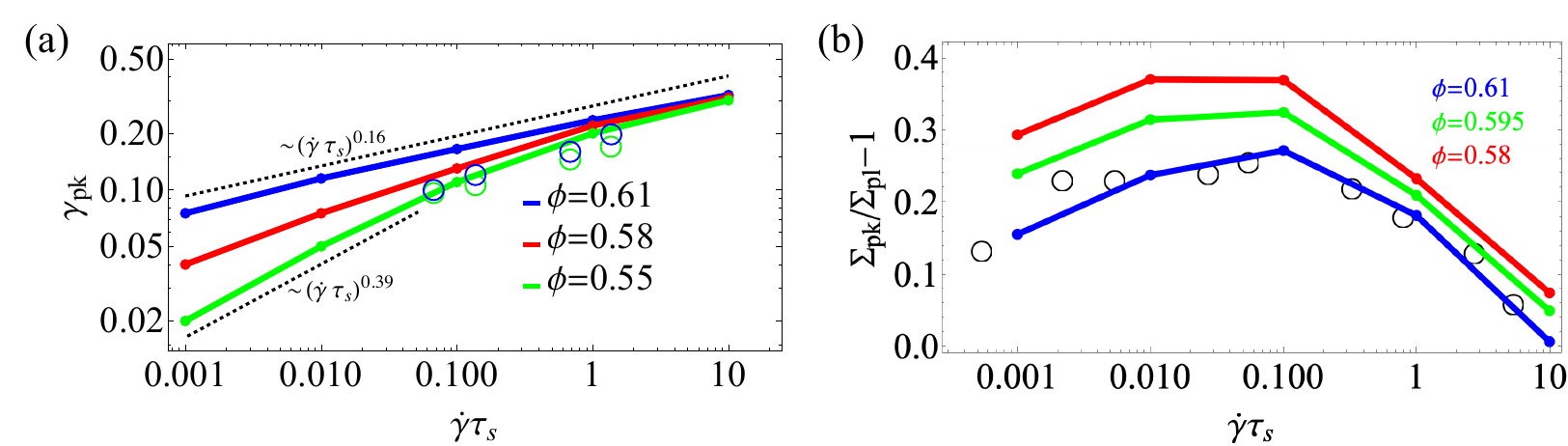}
    \caption{Overshoot properties, theory and experiment. (a) Double log plot of the overshoot strain vs short time Peclet number [59] for different packing fractions. The hollow circles are experimental data for colloids of mean diameter D~535nm at packing fractions of 0.595 (blue) and 0.562 (green)~\cite{koumakis2016a}. (b) Overshoot amplitude as a function of dimensionless shear rate for different packing fractions (see Fig. A1 for the qualitatively different ideal MCT prediction). Hollow black circles are experimental results with D~366nm at $\phi=0.582$~\cite{koumakis2016a}. To compare the shape of the theoretical curves, the experimental points have been vertically shifted upwards by a factor of 3.}
    \label{fig:overshoot_gamma}
\end{figure*}

Fig.10b shows the shear rate dependence of the overshoot amplitude for different packing fractions. Fully aged experimental data for weakly polydisperse hard sphere (HS) suspensions (mean diameter $\sigma=366$nm, $\phi=0.582$) are also shown~\cite{koumakis2016a}. 
A striking non-monotonic dependence is experimentally observed, with $r_O$ becoming very small at both low and high shear rates.
The theory predicts the same behavior for at low and high packing fractions and shear rate ranges typical of experiments [59].

\begin{figure}
    \centering
    \includegraphics[scale=0.83]{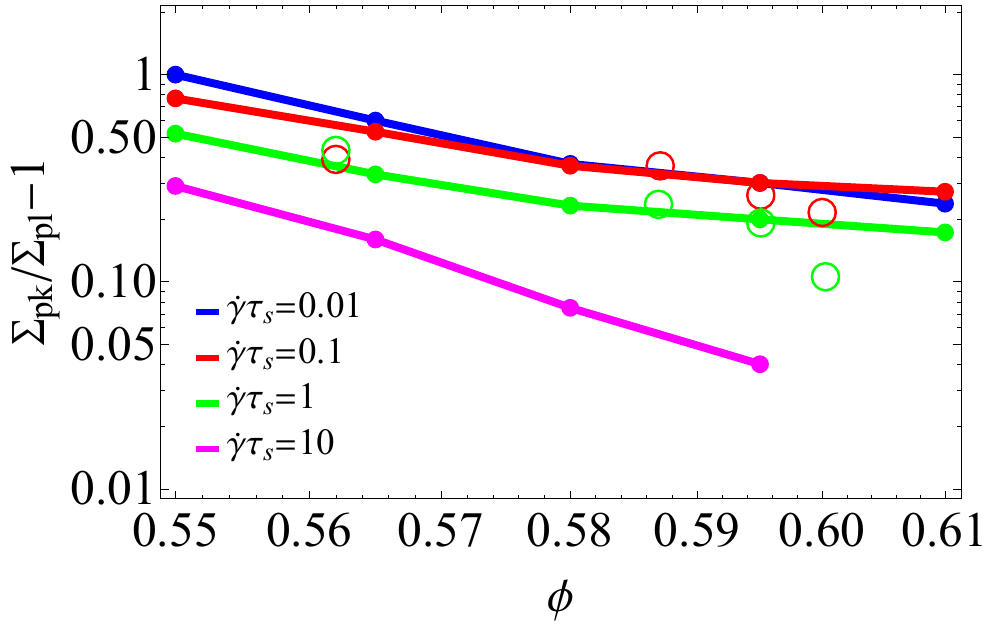}
    \caption{Overshoot amplitude as a function of packing fraction for several shear rates. The experimental data is from the same system as in Fig. 10(b) and is shown as hollow circles for $\dot{\gamma}\tau_s\sim 0.1$ (red) and $\dot{\gamma}\tau_s\sim 1$ (green), respectively~\cite{koumakis2016a}.}
    \label{fig:overshoot_phi}
\end{figure}

Results for the overshoot amplitude as a function of packing fraction at various shear rates are presented in Fig.11. For a fixed shear rate, one sees that $r_O$: (a) decreases monotonically and weakly with packing fraction with a functional form difficult to unambiguously extract, (b) is modestly larger at lower shear rates, and (c) becomes quite small in an absolute sense at the highest shear rates. 
These trends are all in good qualitative accord with the theoretical predictions.  
Considering the challenge of making high precision measurements of the overshoot features, and complexities in real suspensions not in the theory (e.g., particle size polydispersity, hydrodynamic lubrication forces), we find the overall agreement with theory to be very good.

\subsection{Steady State Comparison}

Based on HB form analysis of our numerical theoretical results like those shown in Fig.9, the extracted theoretical “static yield stress” ($\Sigma_{\text{ss}}^{\text{HB}}$) as a function of packing fraction scaled by its RCP value (0.64 for the monodisperse theory, 0.67 for polydisperse experiments~\cite{koumakis2016a,koumakis2012a}) is presented in Fig.12. 
Structural changes under deformation again have little effect, and the predicted nearly exponential form, and tendency to bend over at the highest packing fractions, is in good agreement with experiments on dense hard colloid suspensions~\cite{besseling2010a}\footnote{Using the deformation-dependent structure model, a rescaling constant of ~15 produces quantitative agreement with experimental data. A detailed discussion of the possible reasons for this factor are given in~\cite{ghosh2020a}.}. 
Quantitatively, the theoretical yield stresses are roughly a decade smaller in absolute magnitude than observed experimentally. 
This could have many possible origins given the simplicity of the model and theoretical approximations adopted, including especially how to relate force and stress via an effective cross-sectional area, $A$, as discussed in section II.B.

\begin{figure}
    \centering
    \includegraphics[scale=0.68]{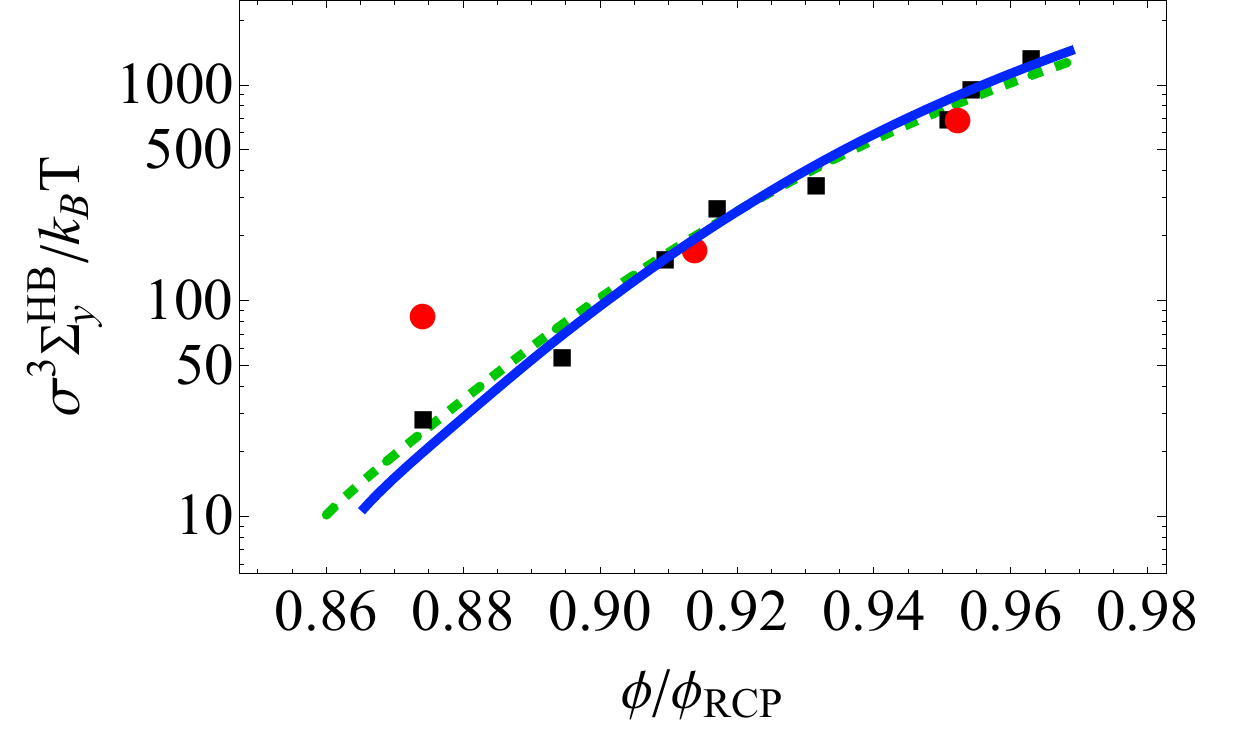}
    \caption{Comparison between the experimentally extracted (with HB model fit) static yield stress~\cite{besseling2010a} with its theoretical analog based on the invariant structure model (dashed green curve shifted up by a factor of 12) and the new theory with $S(k,\gamma)$(solid blue curve shifted up by a factor of 15).}
    \label{fig:hb_expt}
\end{figure}

Recent experimental studies~\cite{cao2021a} of dense emulsions extracted via a HB fit the static yield stress as a function of packing fraction. They found $\sigma_y \sim e^{C\phi}$, where $C \sim 60-70$. 
Our previous analysis in ref.~\cite{ghosh2020a} based on the invariant under deformation structure model predicts the so-called absolute yield stress increases as $\sim e^{60\phi}$ for packing fraction ranges of 0.58-0.61. 
Since incorporating deformation-induced structural changes has nominal consequences on the steady state predictions, our prior conclusion for the dependence of yield stress remains unchanged and is in reasonable agreement with the emulsion results~\cite{cao2021a} within the caveat they employ soft repulsive particles and not strictly hard spheres. 

Based on the above, we conclude that deformation-induced structural changes are critical for understanding the physics of the stress overshoot, but have only minor consequences for the packing fraction and shear rate dependences of steady state properties. 
This deduction is buttressed by our analytic second-derivative analysis of the stress-strain curve presented above. 
Finally, we find that activated dynamics remains important in the nonequilibrium steady state in the sense that activation barriers are nonzero and, most importantly, vary in a systematic manner with packing fraction and shear rate. 
The latter aspect is the origin of our ability to understand the experimental trends discussed above. 
Moreover, as seen in the inset of Fig.8, we note that the barrier height in the steady state relative to its value at the transient overshoot is nearly a universal constant (0.5), independent of shear rate and packing fraction. 

\section{Comments on Other Approaches for Predicting the Transient Shear Response}

The stress-strain curves for a variety of dense (glassy) colloidal suspensions of repulsive particles has modeled by multiple other, generally phenomenological, approaches. 
In this section we briefly mention a few prominent examples, and contrast them with our work. 
We note that essentially all phenomenological models build in by hand the idea that activated processes are important, albeit not at a fundamental microscopic level that connects in a predictive manner interactions, structure, dynamics, and rheology as we have.

Ideal MCT is the technically most directly related approach to our work on glassy hard spheres in the sense that it also is a microscopic force-based theory that relates structure to dynamics. 
A rheological theory has been developed based on an integration through transients formalism where the MCT vertex incorporating force correlations becomes a function of strain, strain rate and time~\cite{fuchs2002a,amann2013a,fuchs2002b,amann2015a}. 
However, since ideal MCT is fundamentally built on the hypothetical idea of a literal kinetic arrest glass transition and includes no activated processes, it can be applied in practice only over a very narrow range of packing fractions defined by the “distance” to the quiescent ideal glass transition (at $\phi_c\sim 0.516$), $\Delta \phi\equiv \phi-\phi_c$. 
Implementation of the full MCT including microscopic structural information for 3d dense suspensions is extremely demanding numerically. 
These considerations have strongly limited confrontation of the predictions of the full ideal MCT with experiment in a comprehensive and no fit parameter manner. 
The Appendix provides a more detailed discussion of the ideal MCT results germane to our work, including the large qualitative differences compared to our theory.

Phenomenological “soft glass rheology” (SGR) models primarily aim to understand the rheological response of systems such as emulsions and slurries composed of soft particles~\cite{sollich1997a,sollich1998a,fielding2000a,fielding2009a}. 
Motivated by phenomenological trap models, deformation and flow is built into local ‘mesoscopic’ zones that can dynamically evolve, and the role of (effectively quenched) structural disorder in determining macroscopic rheological response and metastability is studied. 
Such models can predict a stress overshoot that increases monotonically with strain rate, which can rationalize data on soft pastes but disagrees with experimental observations on hard sphere suspensions~\cite{sollich1998a}. 
Since SGR models are not based on microscopic control parameters, nor directly relate structure, dynamics and rheology, they cannot predict the precise packing fraction dependences of the key rheological features we have focused on in this article.

Phenomenological “elastoplastic models” were originally constructed to explain the stress overshoot of bulk metallic glasses that share some qualitative resemblances to hard-sphere systems. 
Based on the idea of shear transformation zones carrying information of plastic flow, a nonaffine response is coupled to structural rearrangements at the cage level~\cite{zaccone2014b,maestro2017a}. 
While this approach requires at least two non-trivial fitting parameters to characterize relaxation dynamics, the predictions agree qualitatively with some measurements on metallic glasses. 
However, this approach predicts monotonic changes of the stress overshoot magnitude with strain rate, and is unable to capture the packing fraction dependence of this quantity for hard sphere suspensions. 

A ``continuum fluidity'' model has been very recently implemented in detail by combining ideas of the nucleation and growth dynamics of a ‘shear zone’ that acts like a localized fluid state~\cite{bocquet2009a,benzi2019a}. 
The flow properties are controlled by a free-energy functional that depends on a phenomenological (and non-microscopically defined) local fluidity variable, a length-scale related to the size of a fluidized region, and a parameter that carries information about steady-state HB rheology which is adopted empirically. 
A dynamical evolution equation for the fluidity and subsequent minimization of the free-energy gives an evolution equation for stress. 
Significant progress has been made recently with this approach for the nonlinear rheology of soft particle suspensions and pastes~\cite{langer2006a,benzi2021a}. 
However, since the information about the HB flow curve is built into the model by hand, predictions are subject to parameter fitting and a microscopic understanding of how packing fraction changes the rheological response in hard sphere suspensions is not possible. 

\section{Conclusions and Future Directions}

We have constructed a microscopic theory that relates particle interactions, structure, slow dynamics, and rheology based on deformation-assisted thermally activated relaxation that treats in a unified manner the transient and steady state features of the continuous shear rheology of ultra-dense Brownian colloidal suspensions. 
Important physical elements include how activated hopping driven stress relaxation is accelerated due to barrier reduction and the critical role of deformation-induced disordering of local structure as the origin of the stress overshoot. 
The theory captures the rich experimental behavior of dense hard sphere colloidal suspensions nearly quantitatively. 
We emphasize the importance of activated dynamics which is key to our successful predictions of the packing fraction and strain rate dependences of the stress overshoot.  
As shown in the present article and ref.~\cite{ghosh2020a}, the collective elastic physics of central importance for the equilibrium dynamics in highly metastable states plays an essentially negligible role in determining both the overshoot and steady-state rheological behavior. 
This is the physical reason that the prior constant structure model version of the ECNLE theory based rheological approach predicted~\cite{ghosh2020a} a suppression of dynamic heterogeneity with applied strain. 
These prior conclusions remain true in the presence of deformation induced changes of structure both for the transient and steady state responses. 

The magnitude and qualitative aspects of the packing fraction and strain rate dependences of the overshoot and steady state stresses strongly depend on the interparticle interaction. 
Soft microgels, core-shell particles, star-like micelles, and thermoresponsive particles generally show different and sometimes stronger overshoot effects~\cite{koumakis2012b,carrier2009a,pham2008a}. 
The theory presented here can potentially be extended to such systems by employing an appropriate interparticle interaction pair potential and/or intraparticle degrees of freedom. 
For example, for classic microgels, a Hertzian contact potential is often adopted~\cite{ghosh2019a}. 
We believe our same basic theoretical framework can also be generalized to address the rheology of other hard nanoparticle and colloidal dense gels and attractive glasses~\cite{ghosh2019b}, including potentially those that display a “double yielding” behavior~\cite{pham2008a} associated with caging and physical bonding. 

Finally, our theoretical approach has invoked simplifications that would be valuable to address in the future. 
For example, use of an effectively isotropic description of structure and particle trajectories. Addressing this aspect in a microscopic force-level theory will be very complex and computationally difficult, not only because of the explicit vectorial dependence of $S(k)$, particle displacements and dynamic properties, but also because tractable and accurate theories for $S(\vec{k},\gamma)$ largely do not exist. 
A second challenge is to formulate coupled self-consistent evolution equations for $S(k,\gamma)$ and $\Sigma(\gamma)$. 
We note that this problem was addressed in prior work on polymer glasses in the presence of physical aging and mechanical rejuvenation based on NLE theory and the generalized Maxwell description for a highly coarse grained description of polymer chain structure and $S(k)$~\cite{chen2011theory}. 
Ideas in that work may be relevant for constructing in the future a fully microscopic theory of nonequilibrium structure and stress evolution in dense particle suspensions. 
Other real-world complications such as (many body) hydrodynamic interactions~\cite{nabizadeh2021a,nakayama2005a,whittle1997a}, colloids with rough surfaces and dissipative aspects like sliding and rolling friction~\cite{singh2020a,fuchs2014a}, and the shear thickening and/or jamming regimes~\cite{fernandez2013a,wyart2014a} remain outstanding challenges for microscopic nonequilibrium statistical mechanical approaches.

\section*{Acknowledgements}

The early stages of this research was supported by the U.S. Department of Energy, Office of Basic Energy Sciences, Division of Materials Sciences and Engineering under Award No. DE-FG02-07ER46471 through the Materials Research Laboratory at the University of Illinois at Urbana-Champaign. 
The latter stages of the research was supported by the Army Research Office via a MURI grant with contract W911NF-21-0146.

\appendix

\section*{Appendix: Ideal Mode Coupling Theory Predictions}

\begin{figure}[h]
    \centering
    \includegraphics[scale=0.75]{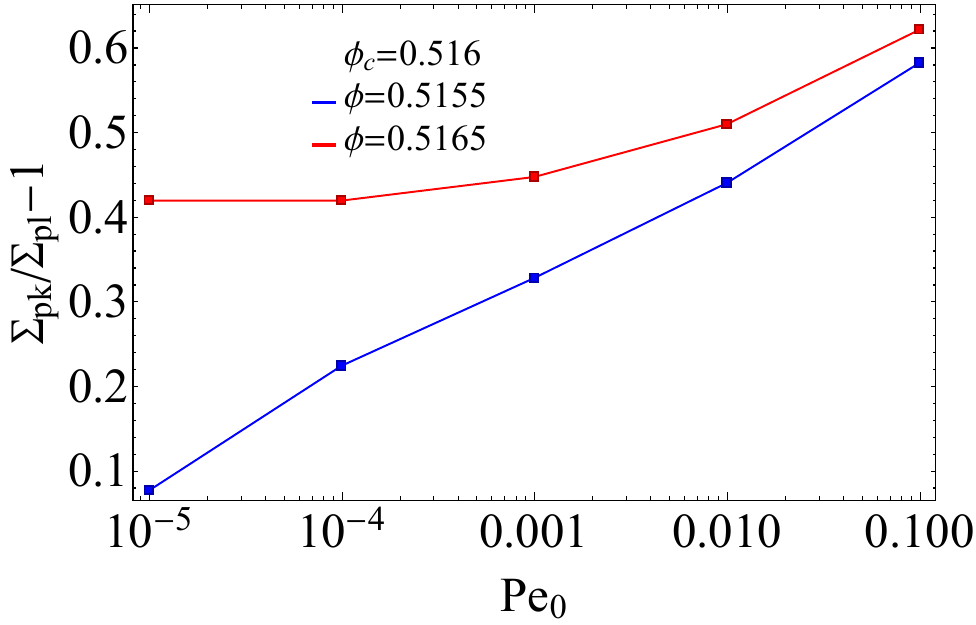}
    \caption{Ideal MCT theory predictions for the overshoot amplitude for monodisperse 3D hard spheres as a function of bare Peclet number for two different packing fractions straddling the quiescent ideal glass transition [30].}
    \label{fig:mct}
\end{figure}

Ideal Mode Coupling Theory (MCT) [31] based approaches for the nonlinear rheology of dense colloidal suspensions [21, 27-30] have largely considered only deformation-induced changes of the static structure factor, $S(k)$, via a simple shear advection mechanism (per a time-dependent wavevector, $k(t)$). 
This is argued to be the crucial process that weakens caging constraints, eventually leading to complete cage ‘melting’ of an “ideal glass” and deformation driven flow as $t\to \infty$. 
Ideal MCT is built on a hypothetical ‘ideal glass transition’ that for hard spheres under quiescent conditions occurs at a critical packing fraction $\phi_c\approx 0.516$ [31] based on structural input from PY integral equation theory.
These defining features have restricted applications of ideal MCT to an exceptionally small range of packing fractions $\phi\in (0.51-0.52)$, significantly below the experimentally studied range of $\phi\sim 0.56-0.61$ or higher where activated dynamics is important in equilibrium. 
Computational complexity has rendered MCT difficult to implement with the full microscopic structural information retained. 
Thus, most prior MCT-based analyses of experimental stress-strain curves and the overshoot have been carried out very close to $\phi_c=0.516$, and are often performed in the context of the further simplified ``schematic'' $F_{12}^{(\dot{\gamma})}$-model [21] and related semi-microscopic models [21, 28-30] which typically have multiple fit parameters. 

Schematic MCT calculations predict that over a very large $\sim 8$ decade change in shear rate the overshoot strain changes only slightly from $\gamma_{\text{pk}}\sim 0.25$ to 0.30[28-30]. 
This is a much weaker variation than observed in experiments and simulations [16-18], and much weaker than predicted by our theory. 
MCT also predicts more than a factor of $\sim 2$ growth of the overshoot amplitude over a tiny change of packing fraction (and a roughly linear growth law with packing fraction) at all fixed shear rates [28-30]. 
This is much too strong and does not agree with simulations or experiments [16-18] nor our theory which find: (i) much weaker changes of the overshoot amplitude, and (ii) an opposite trend with increasing packing at a fixed shear rate, i.e., the overshoot becomes smaller for ultra-dense systems. 
One could in principle calculate the peak strain and magnitude of the stress overshoot from the full microscopic MCT (e.g., using the simplified isotropically sheared hard sphere model [30]) that avoids the schematic aspects and most, but not all, fitting parameters. 
However, results based on this more computationally intensive implementation also lead to a much too large overshoot strain and amplitude with a similar highly limited range of applicability tightly straddling the hypothetical ideal glass transition packing fraction [26-30].

Figure 13 shows a specific example for the overshoot amplitude predicted by the full numerical version of MCT for 3D monodisperse hard spheres [30]. 
The results are dramatically different compared to our theory and the experimental/simulation findings discussed in the main text. 
Ideal MCT results for the overshoot amplitude are shown as a function of the bare $\text{Pe}_0$ for two packing fractions that differ by $\sim 0.001$ on the two sides of the quiescent ideal glass transition at a packing fraction of 0.516. 
As noted above, non-monotonic trends are not predicted, and as the shear rate increases, MCT continues to predict larger values of the overshoot magnitude.

\bibliography{aipsamp}

\end{document}